\documentclass[10pt,journal,compsoc]{IEEEtran}
\IEEEoverridecommandlockouts
\usepackage{color}
\usepackage{xcolor}
\usepackage{multirow}
\usepackage{ifthen}
\usepackage{url}
\usepackage{amsmath}
\usepackage{hyphenat}
\usepackage{amssymb}
\usepackage{xspace}
\usepackage{enumitem}
\setlist{leftmargin=2mm}
\usepackage[ruled,linesnumbered]{algorithm2e}
\usepackage{array,multirow,graphicx}
\usepackage{float}
\usepackage{balance}
\usepackage{tikz}
\usepackage{calc}
\usepackage[caption=false]{subfig}
\usepackage{listings,amsfonts}
\usepackage{amsmath,pifont}
\usepackage{epsfig}
\usepackage{endnotes}
\usepackage{makecell}
\usepackage[normalem]{ulem}
\usepackage{pgfplots}
\usepackage{capt-of}
\usepackage{tcolorbox}
\usepackage{adjustbox}
\usepackage{amsthm}
\usepackage{lipsum}
\usepackage{soul}
\soulregister{\cite}7
\soulregister{\citep}7
\soulregister{\citet}7
\soulregister{\ref}7 
\soulregister{\pageref}7
\soulregister{\tool}{7}
\soulregister{\item}7
\soulregister{\subsection}7
\usepackage{listings}

\definecolor{KWColor}{rgb}{0.37,0.08,0.25}
\definecolor{CommentColor}{rgb}{0.133,0.545,0.133}
\definecolor{StringColor}{rgb}{0,0.126,0.941}
\usepackage{setspace}

\usepackage{cite}
\usepackage{amsmath,amssymb,amsfonts}
\usepackage{algorithmic}
\usepackage{graphicx}
\usepackage{textcomp}

\pgfplotsset{width=7cm,compat=1.16}
\usepgfplotslibrary{statistics}
\graphicspath{ {./} }

\definecolor{yellow}{RGB}{255,255,153}
\definecolor{grey}{RGB}{224,224,224}
\definecolor{green}{RGB}{0,100,0}

\newboolean{showcomments}
\setboolean{showcomments}{true}
\ifthenelse{\boolean{showcomments}}
 { \newcommand{\mynote}[2]{
      \fbox{\bfseries\sffamily\scriptsize#1}
        {\small$\blacktriangleright$\textsf{\emph{#2}}$\blacktriangleleft$}}}
        { \newcommand{\mynote}[2]{}}

\definecolor{DarkOrange}{rgb}{0.8,0.3,0.0} 
\definecolor{DarkCyan}{rgb}{0.0, 0.55, 0.55}
\definecolor{codegreen}{rgb}{0,0.6,0}
\definecolor{codegray}{rgb}{0.5,0.5,0.5}
\definecolor{codepurple}{rgb}{0.58,0,0.82}
\definecolor{backcolour}{rgb}{0.95,0.95,0.92}

\newcommand{\tool}[1]{\textsc{LazyCow}}

\newcommand{\mf}[1]{{\color{cyan}MF: #1}}

\AtBeginDocument{%
  \providecommand\BibTeX{{%
    \normalfont B\kern-0.5em{\scshape i\kern-0.25em b}\kern-0.8em\TeX}}}

\usepackage{url}

\begin{document}

\title{Taming Android Fragmentation through Lightweight Crowdsourced Testing}

\author{Xiaoyu~Sun, Xiao~Chen, Yonghui~Liu, John~Grundy and Li~Li
\IEEEcompsocitemizethanks{
\IEEEcompsocthanksitem Xiaoyu Sun is with the School of Computing, Australian National University, Canberra, Australia. \protect\\
E-mail:  Xiaoyu.Sun.IEEE@gmail.com
\IEEEcompsocthanksitem Xiao Chen, Yonghui Liu, and John Grundy are with the Faculty of Information Technology, Monash University, Melbourne, Australia. \protect\\
E-mail: Xiao.Chen@monash.edu,\\ yonghui.liu@monash.edu and John.Grundy@monash.edu
\IEEEcompsocthanksitem Li Li is with the School of Software, Beihang University, Beijing, China. \protect\\
E-mail: lilicoding@ieee.org
\IEEEcompsocthanksitem Xiaoyu Sun is the corresponding author.
}
\thanks{Manuscript received ; revised.}}

\markboth{IEEE TRANSACTIONS ON SOFTWARE ENGINEERING,~Vol. , No. ,}%
{Shell \MakeLowercase{\textit{et al.}}}

\IEEEtitleabstractindextext{
\begin{abstract}
Android fragmentation refers to the overwhelming diversity of Android devices and OS versions. These lead to the impossibility of testing an app on every supported device, leaving a number of compatibility bugs scattered in the community and thereby resulting in poor user experiences. To mitigate this, our fellow researchers have designed various works to automatically detect such compatibility issues. However, the current state-of-the-art tools can only be used to detect specific kinds of compatibility issues (i.e., compatibility issues caused by API signature evolution), i.e., many other essential types of compatibility issues are still unrevealed. For example, customized OS versions on real devices and semantic changes of OS could lead to serious compatibility issues, which are non-trivial to be detected statically. To this end, we propose a novel, lightweight, crowdsourced testing approach, \tool{}, to fill this research gap and enable the possibility of taming Android fragmentation through crowdsourced efforts. Specifically, crowdsourced testing is an emerging alternative to conventional mobile testing mechanisms that allow developers to test their products on real devices to pinpoint platform-specific issues.  
Experimental results on thousands of test cases on real-world Android devices show that \tool{} is effective in automatically identifying and verifying API-induced compatibility issues. Also, after investigating the user experience through qualitative metrics, users' satisfaction provides strong evidence that \tool{} is useful and welcome in practice.
\end{abstract}

\begin{IEEEkeywords}
Software testing, Crowd-based software engineering, Android Fragmentation, Compatibility Issues.
\end{IEEEkeywords}
}

\maketitle
\IEEEdisplaynontitleabstractindextext
\IEEEpeerreviewmaketitle

\section{Introduction}
\label{sec:introduction}

Fragmentation has long been a severe issue in Android, causing many compatibility issues that may make apps crash on users' Android devices and subsequently lead to poor user experiences.
Indeed, there are a massive number of Android OS versions and customized ROMs (i.e., customized Android firmware by smartphone manufacturers) available in the market. 
The heavy fragmentation issues make it hard for Android app developers to carefully test their apps w.r.t. compatibility issues across the many different Android devices. 
According to Joorabchi et al.\cite{joorabchi2013real}, Android device fragmentation is a big challenge for development as well as for testing. 76\% of their survey participants (Android mobile app developers) see fragmentation as one of the most challenging tasks because they have to test their apps on different OS versions and screen sizes to ensure that their app works.

Given the fact that Android is an open-sourced framework, which enables device vendors and OS providers to customize OS versions, it results in a large variety of customized devices on the market~\cite{theverge_android}. Unfortunately, the exponential growth of customized Android frameworks has led to serious compatibility issues in the Android ecosystem, as recently shown~\cite{liu2022customized, cai2019large, he2018understanding, scalabrino2019data, wei2016taming, xia2020android}. For example, Liu et al.~\cite{liu2022customized} demonstrate that the textual merge conflicts may introduce compatibility issues to customized OS versions. Cai et al.~\cite{cai2019large} further experimentally claim that the diversification of Android devices is one of the primary causes of incompatibility issues in Android. As a consequence, the productivity of app developers can be heavily impeded as they have to test functions on as many devices as possible to ensure no incompatibility issues that may cause poor user experiences. Theoretically, developers should physically collect devices with different brands, models, SDK versions, software/hardware configurations, etc. However, having dedicated devices covering all specifications for each app developer is not practical. In addition, it is time-consuming for app developers to integrate the process of incompatibility testing into their daily continuous integration workflow. Therefore, we argue that there is a need to tame Android fragmentation issues in a lightweight crowdsourced manner.

To the best of our knowledge, at the moment, state-of-the-art works detect compatibility issues mainly through static analysis techniques~\cite{ham2011mobile, huang2018understanding, li2018cid,wei2016taming,zhang2015compatibility}. For example, Li et al.~\cite{li2018cid} present a static approach called CiD that models the evolution of Android APIs and inspects app bytecode to detect the misuses that may cause incompatibility issues. However, static approaches are known to suffer from a high false positive rate as it has to accept trade-offs to obtain relatively good results. In addition, static techniques may be susceptible to sophisticated programming features (e.g., reflection, obfuscation, and hardening)~\cite{sun2021taming}, harming the soundness of these approaches. Furthermore, static techniques are reported to only be effective in detecting specific types of compatibility issues~\cite{liu2021identifying} (i.e., compatibility issues caused by syntactic changes), leading to other complicated types of compatibility issues uncovered. For example, according to Sun et al.~\cite{sun2022mining}, CiD is unable to handle compatibility issues that are triggered by semantic changes. Moreover, compatibility issues could also be introduced by the customization of Android OS, which is non-trivial to be detected statically. To mitigate this problem, we propose a lightweight crowdsourced platform to automatically distribute tests across real-world devices so that we are able to detect a wider range of compatibility issues by taking the advantage of dynamic 
testing. 


In this work, we present a lightweight crowdsourced testing framework, \tool{}, that attempts to automatically distribute and run collected test cases on real-world devices to trigger compatibility issues dynamically. Being ``lightweight", it means that we want to directly dispatch and execute relevant test cases to crowdsourced devices, instead of dispatching executable Android apps as is usually done in the state-of-the-art approaches.
Compared with traditional crowdsourced app testing, we believe our lightweight approach can bring the following key benefits:
(1) Reducing bandwidth so that will only have minimal impact on users' everyday activities when using the phone.
(2) Diminishing users' awareness about the crowdsourced testing as the testing app only needs to install once (i.e., the first time, after that, only test cases will be dispatched). 
(3) Allowing flexibility and hence can easily achieve on-demand testing and continuous testing. In \tool{}, only a certain number of test cases will be dispatched and executed based on the smartphone's running status.
New test cases can also be continuously loaded and executed when the smartphone becomes idle for a period of time.
(4) Guarantee full test case execution as each test case is specifically designed to test the target APIs (e.g., it does not involve complicated logic to reach the APIs). This is unlike testing a whole app, where there is often a challenge to cover all the app code under testing.

To the best of our knowledge, no approaches have yet been developed that use the concept of lightweight crowdsourced app testing.
We propose in this work a novel lightweight crowdsourced testing approach, \tool{}, to fill this gap and to enable the possibility of taming Android fragmentation through crowdsourced efforts.
To do this we designed and implemented a prototype tool called \tool{}, which leverages a client-server platform to achieve the purpose, of distributing test cases to crowd-sourced smartphones for checking potential compatibility issues. The \tool{} client app needs to be installed on Android users' devices.
It aims to determine a suitable time when the device is in an idle state to (1) interact with the server for requesting (new) test cases, (2) execute the test cases, and (3) send the execution results to the server app for further analysis.
By testing the test cases extracted from the official Android framework and 1,000 random Android apps with \tool{}, we successfully detect 393 APIs that have compatibility issues. After manual validation, we confirm that all detected issues are true positives, suggesting a 100\% true positive rate. Additionally, we find that 109 of them are Signature-based issues, while 284 are Semantics-based issues and cannot be noticed by state-of-the-art static methods. Also, we identify that 161 and 47 compatibility issues are vendor- and model-specific, respectively, which are introduced when smartphone vendors customize the Android system. Such vendor/model-specific compatibility issues may introduce severe security problems.

We make the following key contributions to this work:
\begin{itemize}
\item \textbf{New technique.} We have designed and implemented a lightweight prototype tool, \tool{}, which leverages crowdsourced testing techniques to automatically dispatch and execute test cases on Android devices for identifying  fragmentation-introduced compatibility issues.

\item \textbf{New discoveries.} We have demonstrated the effectiveness of \tool{} by dispatching and executing thousands of test cases on real-world Android devices. The experimental results reveal semantic compatibility issues (including vendor/model specific issues), which are overlooked by state-of-the-art approaches. 

\item \tool{} is demonstrated to be useful in practice based on the positive feedback given by real-world Android users. The high satisfaction score suggests most users are happy with the performance of \tool{} in terms of simplicity, satisfaction and adoption willingness.

\end{itemize}

The source code of both client side\footnote{\url{https://github.com/sunxiaobiu/LazyCow}} and server side \footnote{\url{https://github.com/sunxiaobiu/RemoteTest}} are all made publicly available in our artifact package.

\section{Background and Motivation}

\begin{figure}[!t]
    \centering 
    \includegraphics[width=0.4\textwidth]{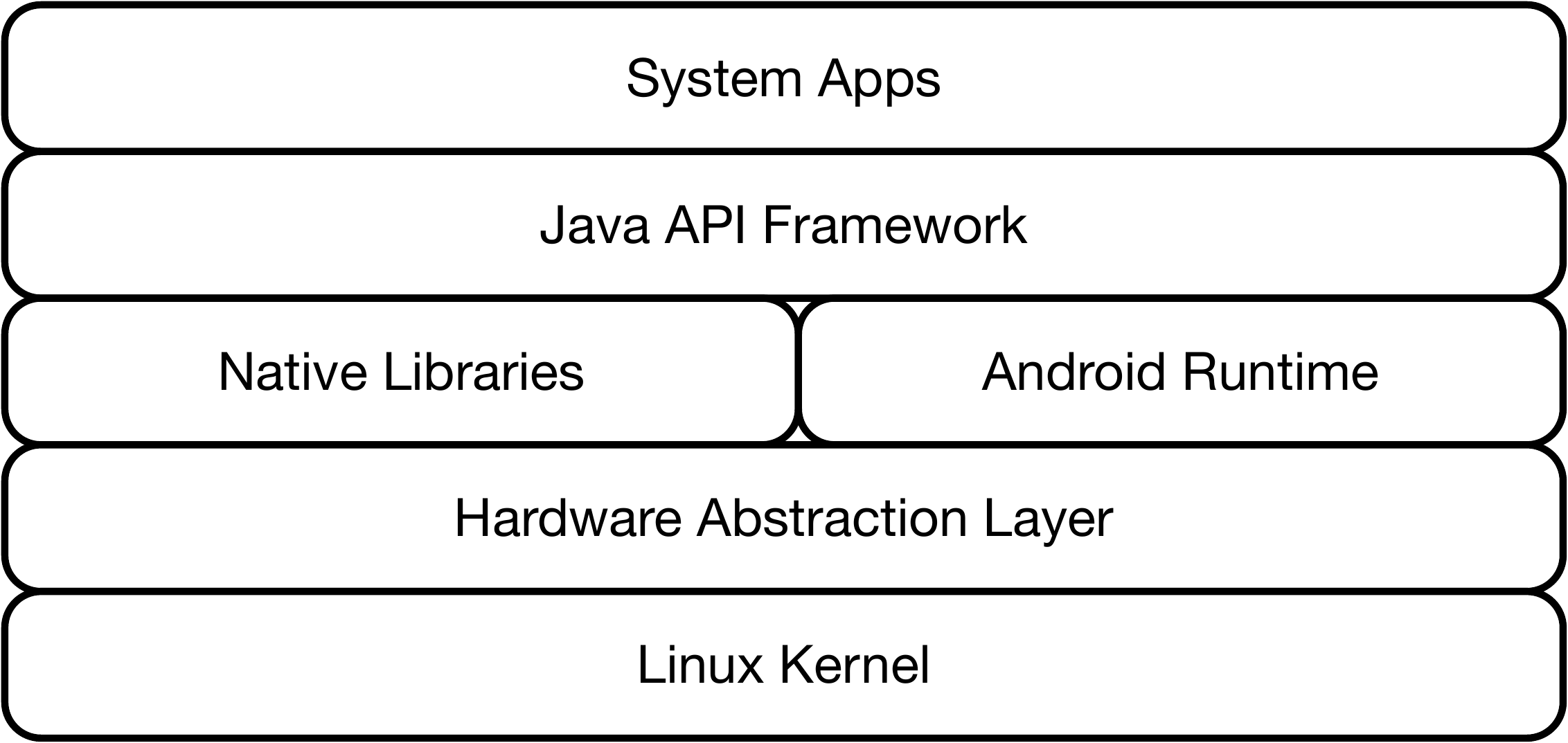}
    \caption{The Android system architecture.}
    \label{fig:android_architecture}
\end{figure}

\subsection{Android Fragmentation and Google CTS} 
Although Android offers device vendors and OS providers an open and flexible software architecture that allows them to swiftly launch their Android device products, it complicates the work of app developers to tolerate sophisticated variability in customized OSs and produce compatible Android applications on diverse device models. Due to various customized Android OS versions, Android fragmentation is becoming a notorious issue, which stems from the necessity to accommodate the proliferation of multiple Android devices with varying software and hardware environments~\cite{lu2016prada}. Specifically, the root causes of fragmentation issues can be summarized into the following two categories:

\begin{itemize}
\item \textbf{The evolving of Android Framework.}
As the Android operating system evolves and releases new versions each year, its API specifications rapidly change syntactically and semantically, leading to compatibility issues~\cite{joorabchi2013real, choudhary2015automated, han2012understanding}. Specifically, a significant proportion of APIs is available only on a few API levels, mainly because they are removed from certain SDK versions or because their implementation changes semantically~\cite{li2018cid}. As a result, devices associated with different OS versions may suffer from  API-level compatibility issues.

\item \textbf{Diverse Device Models.} 
Android device vendors continue to release new device models with various hardware (e.g., varying screen sizes, camera quality, and sensor compositions) and customized OS versions to meet market needs. Specifically, they tend to customize the source code of the Java API framework (as shown in Figure~\ref{fig:android_architecture}) to meet the requirements of different device models. Sometimes, device vendors even programmatically revise the hardware abstraction layer (in Figure~\ref{fig:android_architecture}) to satisfy special needs~\cite{wei2018understanding}. Such customisation is blamed to cause device-specific compatibility issues, which are regarded as non-trivial to be detected in practice.
\end{itemize}

To mitigate the impact of the fragmentation issues, Google, the main contributor of Android, has proposed to the community a Compatibility Test Suite (hereinafter referred to as CTS)~\cite{CTS}, which is a free, commercial-grade test suite.
This test suite contains thousands of test cases targeting different compatibility issues, which may be encountered by given Android apps when running on certain Android devices.

However, CTS is known to be incomplete~\cite{park2013fragmentation} and ineffective in detecting compatibility issues. On the one hand, the test cases included in CTS may not be adequate to help in identifying all the possible compatibility issues. Certain compatibility issues may have been overlooked, especially those caused by OS evolution. Park et al.~\cite{liu2016concurrent} reveal that in practice, fragmentation issues remain in Android despite applying CTS testing. The primary reason is that the criteria for passing the compatibility test suite have not been specified completely. Specifically, CTS only checks for incompatibility among devices, not applications, so in the current status, CTS cannot be utilized in app development.

On the other hand, the CTS is often only applied before the release of Android devices. However, the Android system keeps evolving; hence compatibility issues could be introduced after the devices' release, which has been experimentally confirmed by our fellow researchers ~\cite{li2018cid, wei2016taming, zhou2014peril,wu2013impact}. Moreover, CTS is designed only for device manufacturers who normally execute test suites on the application level. However, compatibility issues are reported to frequently appear at the API level, indicating that the execution of CTS is not sufficient to trigger compatibility issues completely. Furthermore, CTS is practically too heavyweight for app developers to integrate into their version-update workflow. App developers must guarantee that their apps are compatible with a variety of device models and support various OS versions as well. This takes a substantial amount of testing and diagnosis. In practice, it is extremely hard for developers to thoroughly test a piece of code (i.e., unit test) on all possible combinations of device models and Android OS versions.

To this end, we are motivated to provide a lightweight approach to dispatch test cases on real-world devices so as to detect a wider range of compatibility issues.





\subsection{Crowdsourced Android App Testing}
Crowdsourced testing has been a hot research topic for many years~\cite{orso2002gamma,memon2004skoll,elbaum2004empirical}. It has also been applied to the Android field, e.g., to achieve crowdsourced app testing.
For example, on the industry side, the company Global App Testing~\cite{Global_App_Testing}, a leading end-to-end functional testing company, provides access to global testers to execute tests on real devices manually. Such a strategy enables their customers to reduce the time and effort of finding the bugs with the help of professional testers around the world. Other popular crowdsourced testing companies, such as Digivante~\cite{Digivante}, test IO~\cite{test_IO}, and QA Mentor~\cite{QA_Mentor} provide similar crowdsourced testing services, enabling users(e.g., app developers) to test their mobile apps with thousands of professional testers. Unfortunately, all these state-of-the-practice approaches have crowd workers involved, which means none of them can automatically tame Android fragmentation in a lightweight way (i.e., without human intervention). Furthermore, users (e.g., app developers) cannot customize test scripts to meet their particular needs, leading to certain types of incompatible issues undetected. 

On the other hand, crowdsourced app testing has also grown as a trend in academia~\cite{wu2017appcheck, guo2020crowdsourced, li2019cocotest, liang2019summarizing, chen2019automatic, zhang2017crowdsourced}. For example, Wu et al.~\cite{wu2017appcheck} present an approach to record the user interactions and then replay them on devices through a crowdsourced testing service. However, they generate test scripts using the record/replay technique, which is time-consuming as real users' interactions with apps are involved. Li et al.~\cite{li2019cocotest} further develop a platform, CoCoTest, which exploits the concept of collective intelligence to recommend bug reports to the workers. Unfortunately, this approach can be ineffective since crowd workers tend to submit low-quality bug reports.

To sum up, all the state-of-the-art crowdsourced testing platforms involve human intervention, which is time-consuming and error-prone (human intervention can be heavily affected by different levels of professionalism).
Generally speaking, crowdsourced app testing has mainly been applied to test the whole Android app on crowdsourced devices.
This approach shares the drawback of dynamic app testing, i.e., it cannot easily cover (hence explore) all the app code. Our community has never explored the possibility of distributing test cases (i.e., directly executable code snippets) to real-world Android devices.
This evidence motivates us to present a platform to automatically generate and distribute tests on real-world devices without human intervention to automatically detect Android compatibility issues.

\section{Our Approach}
Motivated by the above-mentioned drawbacks of CTS and existing crowdsourced testing platforms, we propose a new lightweight crowdsourced dynamic testing approach. We aim to help the community defend against possible compatibility issues caused by the fragmentation problem of the mobile ecosystem.

\textbf{Why lightweight? } First, our approach is lightweight since we apply crowdsourced testing for test cases only. Unlike traditional Android app crowdsourced testing, which often dispatches executable Android apps to global testers, we dispatch single test cases (i.e., each containing only a small piece of code snippets) to detect API-level compatibility issues.
Our lightweight approach saves significant bandwidths when transferring the test objects to client devices, e.g., an Android app can be at least several megabytes (and could be even hundreds of megabytes), while a test case will only be a few bytes.
Furthermore, it is also non-trivial to thoroughly test an Android app, although it is straightforward to trigger a single test case.
Indeed, the state-of-the-art automated app testing approaches are still not mature enough to achieve a high testing coverage. As such, the targeted code in the dispatched app could not even be reached, resulting in not only heavy but also poor crowdsourced testing results. 
Moreover, dispatching a whole app for crowdsourced testing may also increase the possibility of including malicious payloads, which could threaten the security of the users and, in turn, discourage the adoption of this kind of crowdsourced testing approach.

In this work, instead of carrying out testing as a whole integrated package and involving human intervention, we resort to a lightweight crowdsourced testing approach to achieve our objective. Our approach provides a more promising way by using a dispersed strategy to dispatch test cases (rather than whole Android apps) to be directly executed on diverse real-world Android devices. Crowdsourced testing offers Android developers an opportunity to have their customized test cases to be tested by real users on real devices across the globe, ensuring a customer-centric emphasis.

\textbf{Why Dynamic Testing? } While fragmentation has been a severe issue in the Android ecosystem, there have been several works~\cite{li2018cid, wei2019pivot, xia2020android} proposed for taming Android fragmentation issues through static analysis. However, static analysis approaches are known to lack concrete evidence to confirm these issues on real-world devices. It is also difficult for static analysis approaches to observe code's semantic changes, for which the aforementioned approaches have overlooked. For example, APIs with semantics-based compatibility issues (i.e., have the same signature but different implementation) may evolve and contradict with developer's initial expectations.

We now present a concrete example to demonstrate why static methods are insufficient to identify compatibility issues, so as to better motivate our work.
An Android API called \emph{android.app.usage. NetworkStatsManager
\#querySummary} suffers from both signature and semantic compatibility issues. The test case throws \emph{NoClassDefFoundError} on SDK version 21 and 22, throws \emph{SecurityException} on SDK version 24, while successfully executed on SDK version 25 to 30. Through in-depth analysis, we find that the class \emph{NetworkStatsManager} and the API were both first introduced at API level 23, and it requires permission \emph{PACKAGE\_USAGE\_STATS} to access only at API level 23. However, from API level 24, this API no longer requires permission to access, leading to semantic compatibility issues. 
It is non-trivial to examine if there is a compatibility issue after API level 23 by only statically looking at the Java code of the framework. The state-of-the-art static analysis tool CiD~\cite{li2018cid} fails to detect such semantics changes because it only examines the change of API signatures (including name, type, and parameters). As a result, it leads to many false negatives caused by imprecisely extracting the implementations of the APIs. To that end, we resort to a dynamic crowdsourced approach to dispatch and execute the target API on real-world Android devices to examine the compatibility issues based on its runtime behaviours.

\section{Architecture of \tool{}}

\begin{figure}[!h]
    \centering
    \includegraphics[width=1.05\linewidth]{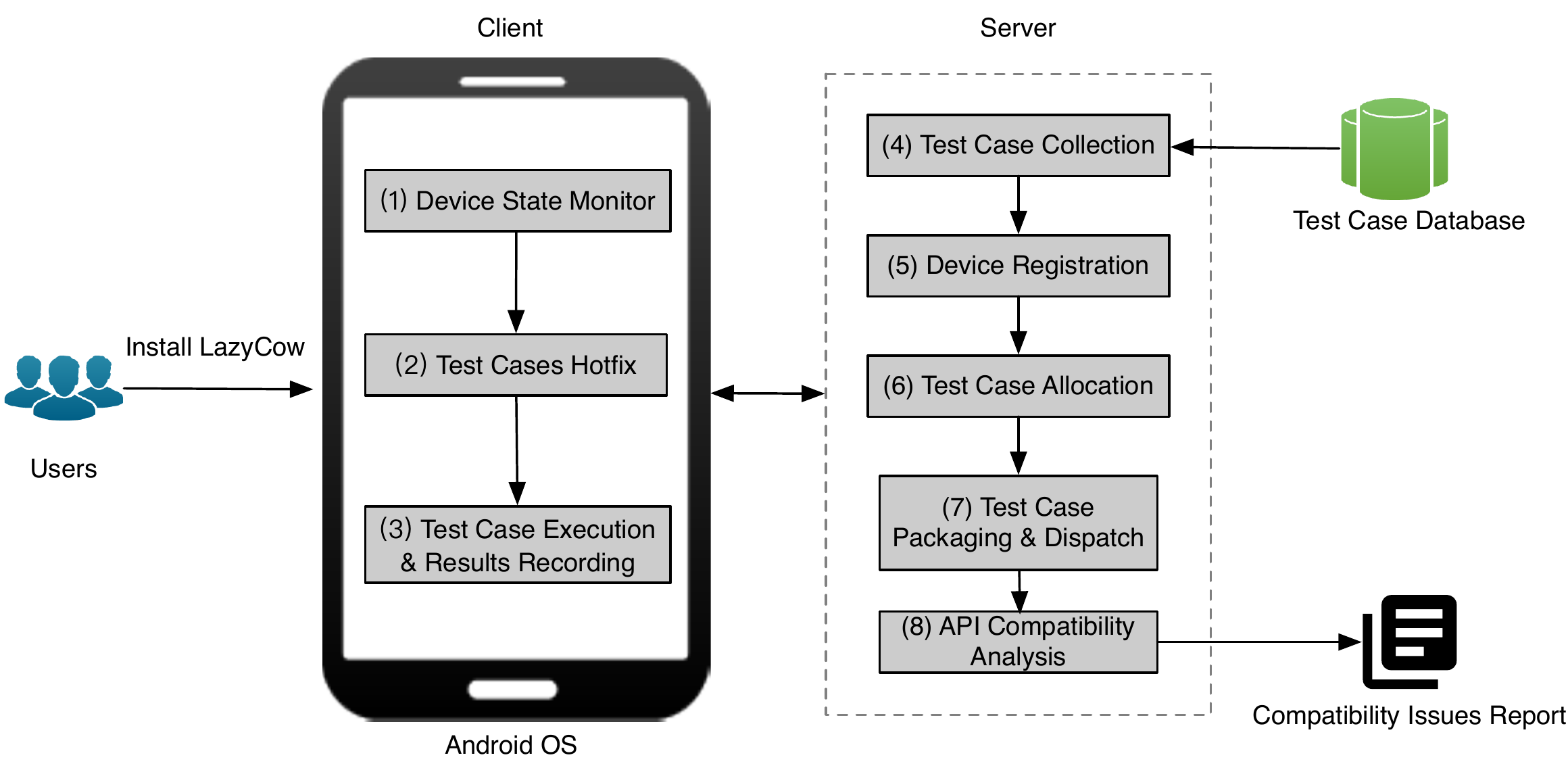}
	\caption{The working process of our approach.}
    \label{fig:LazyCow_architecture}
\end{figure}

The main goal of our work is to provide a lightweight crowdsourced testing platform for automatically dispatching and executing unit tests on various real-world Android devices. To that end, we design and implement a prototype tool called \tool{} for achieving this purpose. Figure~\ref{fig:LazyCow_architecture} illustrates the architecture of \tool{} and it works in a client-server model. The client is installed on various android devices and manages the execution of test cases. Specifically, it determines the time and number of test cases to be executed and sends the testing results back to the server for further analysis.
The server manages the collection of test cases, packaging and dispatching of test cases to the clients, and analyzing the compatibility issues based on the results collected from various devices. We elaborate on the detailed process of each component in the following subsections.


\subsection{Client Side}
We developed a \tool{} client app to be installed on Android devices. The client monitors the status of the device to determine the suitable time (e.g., when the device is not in use) to run the test cases. It then interacts with the server to download and execute the test cases and return the execution results to the server. As shown in Figure~\ref{fig:LazyCow_architecture}, the client consists of three modules, namely \emph{(1) Device State Monitor}, \emph{(2) Test Cases Hotfix}, and \emph{(3) Test Cases Execution \& Results Recording}.

\textbf{(1) Device State Monitor.}
In order to not disturb the users' experience, \tool{} monitors the state of the devices to determine a suitable time to run the test cases. In this work, our definition of a suitable time is the moments that satisfy the following three conditions:
\begin{enumerate}[label=\roman*]
\item \textbf{Phone State:} We use \emph{android.os.PowerManager\#isDeviceId\\leMode} and \emph{android.os.PowerManager\#isScreenOn} to check if the user is interacting with the device. We consider it a suitable time if the user is not interacting with the device.
\item \textbf{Memory Usage:} We use \emph{android.app.ActivityManager\#get\\MemoryInfo} to grasp the memory usage of the device. If it is lower than 25\%, we consider it a suitable time.
\item \textbf{Battery State:} We use \emph{android.os.BatteryManager} to check the battery state to see if it is charging and has sufficient battery life (e.g., above 60\%).
\end{enumerate}

Once a suitable time is detected, the client initiates a request to the server to download the test cases for testing.

\textbf{(2) Test Cases Hotfix.} 
In order to dynamically dispatch incremental tests on Android devices without app reinstalling, we resort to hot-fix technique that supports class, so File and resources updated with the least amount of impact to the user experience. We start by investigating the  state-of-the-practice hotfix solutions~\cite{bao2019android}. Specifically, there are several hotfix tools on the market, the most well-known of which are Tinker, AndFix, Robust, and QZone. To select the most suitable one, we compared their advantages and disadvantages, which are shown in Table~\ref{tab:comparison_hotfix}.

\begin{figure}[!h]
    \centering
    \includegraphics[width=0.6\linewidth]{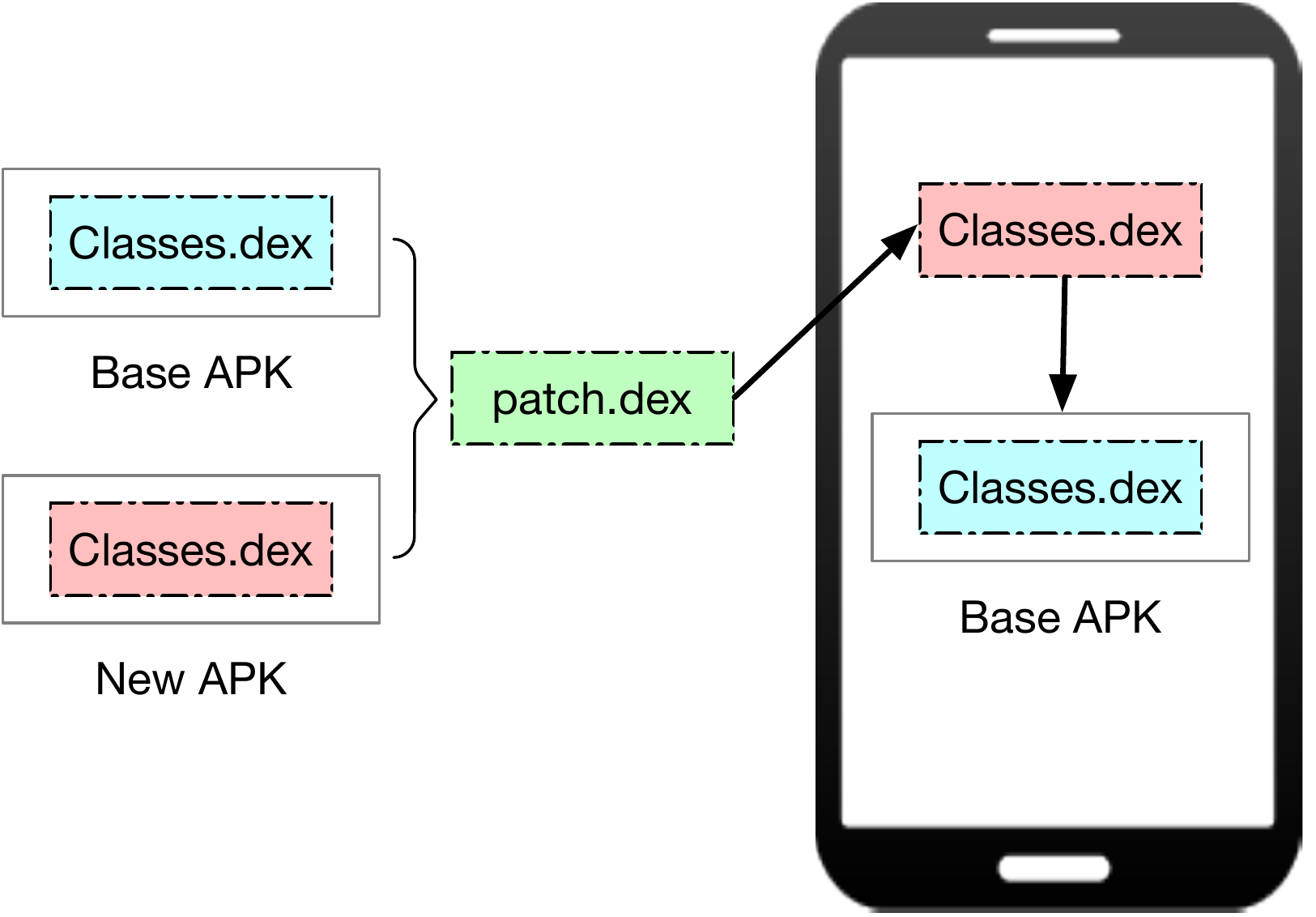}
	\caption{The repair principle of Tinker.}
    \label{fig:Tinker}
\end{figure}

In summary, AndFix and Robust are incapable of doing class/variable replacement, which is what LazyCow mainly asked for. Besides, they do not support the Gradle build tool, which is a necessity for building android applications. In terms of QZone, although it is capable of replacing classes, its greatest disadvantage is the performance problem caused by Dalvik instrumentation. In other words, QZone would significantly increase the memory usage and power consumption of the user's phone. To this end, compared our needs to other hotfix tools, Tinker is the better choice that supports class replacement with smaller performance loss.

\tool{} integrates \emph{Tinker} \cite{tinker}, a hotfix solution when downloading test cases from the server. Tinker supports DEX, libraries, and resource updates without reinstalling the APK, which has minimal influence on the users when updating incremental test cases.
The repair principle of Tinker is based on class loading, and it supports the replacement and addition of classes and resources in terms of functions. The figure~\ref{fig:Tinker} describes the repair principle of Tinker, mainly based on the DEX subcontracting scheme and uses the principle of multiple DEX loading. After comparing the differences between the new APK and the base APK, the updated classes and resources are merged into a file \emph{patch.dex}. Then, the \emph{patch.dex} is combined with the applied classes.dex, and then replace the old DEX file as a whole to achieve the purpose of the hotfix.

According to the idea of replacing the new Dex with the full amount of instant run, we decided to package the test cases into the dex files of an APK. Specifically, during the process of hotfixing, the differences between the old and new test cases are calculated and placed into the patch package, which is then synthesized to the device for hot update. Following that, incremental test cases in the patch package would be put in the directory beneath Tinker. Then, based on the principle of hotfixing, Tinker's Classloader is able to load new test cases in the patch package.

\begin{table}[!h]
\centering
\caption{The comparison results between the state-of-the-practice hotfix tools.}
\label{tab:comparison_hotfix}
\resizebox{0.9\linewidth}{!}{
\begin{tabular}{ | l | l | l | l | l | } 
\hline
 & Tinker & QZone & AndFix & Robust \\
\hline
Class Replacement  & yes & yes & no & no\\
\hline
So File Replacement  & yes & no & no & no\\
\hline
Resource Replacement  & yes & yes & no & no\\
\hline
Full Platform Support  & yes & yes & yes & yes\\
\hline
Effective Immediately & no & no & yes & yes\\
\hline
Performance Loss  & small & large & small & small\\
\hline
Patch Size & small & large & medium & medium\\
\hline
Transparent Development  & yes & yes & no & no\\
\hline
Complexity & low & low & high & high\\
\hline
Gradle Support  & yes & no & no & no\\
\hline
Rom Size & large & small & small & small\\
\hline
Success Rate  & high & high & medium & highest\\
\hline
\end{tabular}}
\end{table}

\textbf{(3) Test Cases Execution \& Results Recording.}
After test cases are downloaded to the client devices, \tool{} leverages \emph{reflection calls} to retrieve all the test cases in the DEX files and execute them sequentially. Since the test cases are written in the format of the Java Unit test, \tool{} automatically runs test cases based on the annotation of each test method. 

JUnit~\cite{JUnit} is the most recommended unit testing framework in Java. There are 5 annotations for test execution callbacks in JUnit:\emph{@BeforeClass}, \emph{@Before}, \emph{@Test},\emph{@After}, \emph{@AfterClass}. Specifically, the test methods are annotated by the \emph{@Test} annotation. In addition, it also supports to constrain the execution flow of certain methods. For instance, to define a method to be executed before (or after) the test methods with the \emph{@Before} (or \emph{@After}) and \emph{@BeforeClass} (or \emph{@AfterClass}) annotations. To that end, \tool{} firstly conducts a static analysis to resolve the annotations from each method. Then, \tool{} leverages reflection call to invoke methods in the sequence of \emph{@BeforeClass} $\rightarrow$ \emph{@Before} $\rightarrow$ \emph{@Test} $\rightarrow$ \emph{@After} $\rightarrow$\emph{@AfterClass}. 

After executing the test cases, \tool{} uses a \emph{try-catch} block to handle the exception that may occur. \tool{} collects the execution result whenever the test case fails or succeeds, with relevant information (e.g., the stack trace information when a test failed) and sends it back to the server for further analysis.


\subsection{Server Side}
The server maintains a test case database that is expected to be collected from various sources (e.g., AOSP codebase \cite{sourcecode_aosp}, Github app code repositories, etc.). These test cases are packaged and dispatched to the registered clients in a load-balancing manner. After finishing executing the test cases on the client devices, the server gathers the outputs for further analysis to identify potential compatibility issues. As demonstrated in Figure \ref{fig:LazyCow_architecture}, the modules on the server side include \emph{(4) Test Cases Collection}, \emph{(5) Device Registration}, \emph{(6) Test Cases Allocation}, \emph{(7) Test Cases Packaging \& Dispatch}, and \emph{(8) API Compatibility Analysis}.

\textbf{(4) Test Case Collection.}
The server maintains and continuously updates a database of test cases to be tested on the client devices. 
The test cases can be collected from the following three sources:
\begin{itemize}
\item The test cases included in the Android Open Source Project (AOSP) codebase~\cite{sourcecode_aosp}, which are written by Android OS developers.
\item The test cases generated by automatic test case generation tools, such as JUnitTestGen~\cite{sun2022mining}, which mines the API usage to generate unit test cases for pinpointing compatibility issues. According to Sun et al.~\cite{sun2022mining}, generic test case generation approaches (such as EvoSuite~\cite{fraser2011evosuite}) are aimed at generating tests for classes (not at the API level), and they have been demonstrated as insufficient in pinpointing compatibility issues because of the lack of API usage knowledge. To this end, given that \tool{} is also tailored for detecting compatibility issues, JUnitTestGen has been selected because it is the more suitable one that outperforms EvoSuite in generating tests for compatibility testing.
\item LazyCow users can also write custom test cases to meet their particular needs. For example, in the daily continuous integration of app development, app developers may want to check if a certain API involves compatibility issues on specific Android devices.
\end{itemize}

\textbf{(5) Device Registration.}
The client registers itself with the server once the \tool{} client app is installed. The device information, such as the manufacturer and model of the device, SDK version, device language, and device screen size, will then be recorded for optimizing the test case dispatch process. Note that \tool{} does not collect personal private data such as device ID, but instead assigns each device a unique ID.

\begin{figure}[!h]
 \setlength{\abovecaptionskip}{0pt}
 \setlength{\belowcaptionskip}{5pt}
    \centering
    \includegraphics[width=\linewidth]{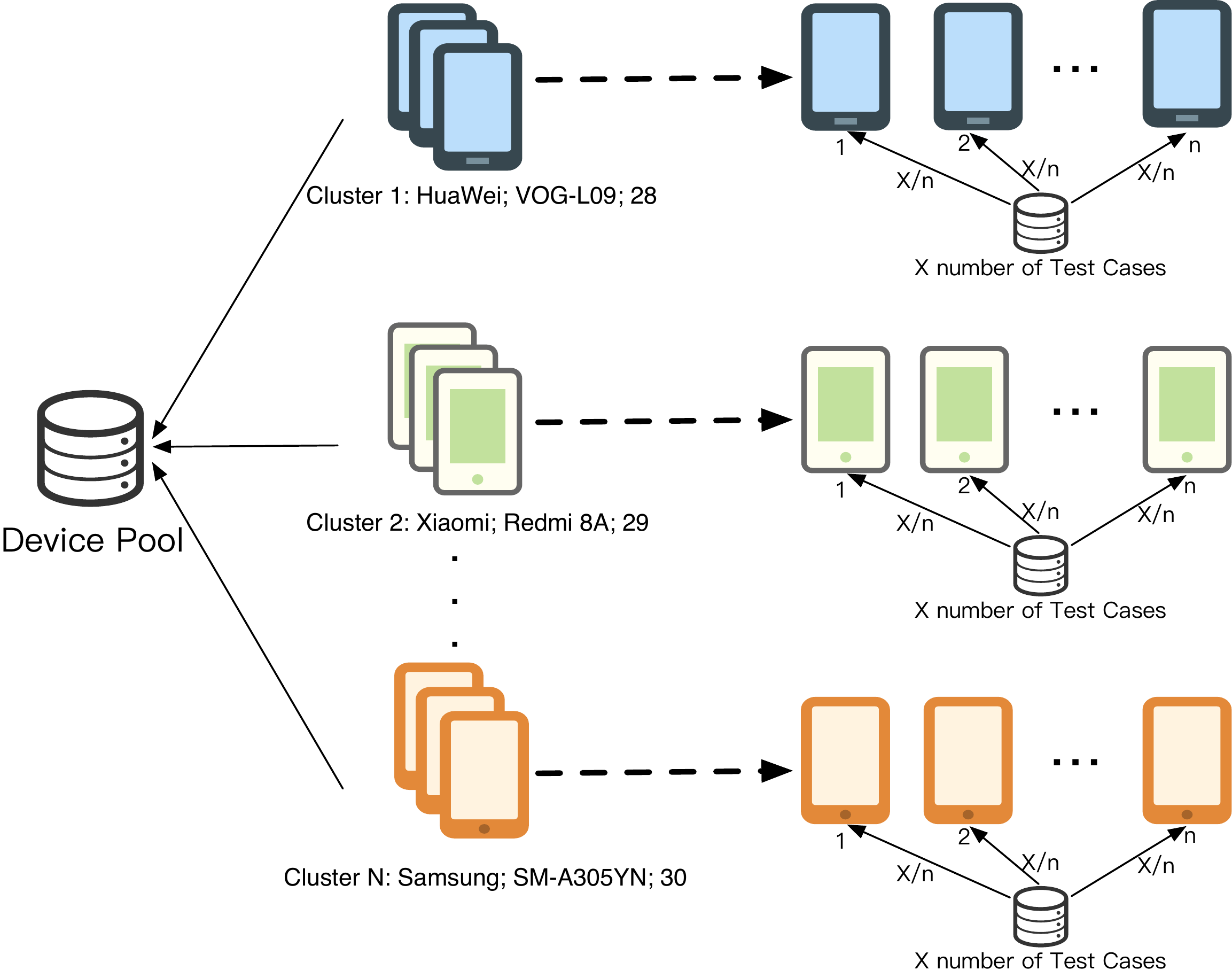}
	\caption{Test case allocation with the load-balancing strategy.}
    \label{fig:TestAllocation}
\end{figure}

\textbf{(6) Test Case Allocation.}
To allocate test cases in a load-balanced way across all available registered devices, we design a test case allocation algorithm to determine the test cases to be assigned to different devices. Figure~\ref{fig:TestAllocation} describes the process of test case allocation with a load-balancing strategy. Specifically, \tool{} first classifies registered devices into device clusters based on the device information, such as the device's manufacturer, the device's model, the Android SDK version, etc (e.g., Android devices with Brand Huawei, Model VOG-L09 and API level 28 would be collected and assigned to the cluster 1). Each cluster will include all devices with the same specifications. The test cases are then equally assigned to all devices in each cluster. By doing so, we make sure that each test case is executed on as many devices with different specifications, and no test cases are redundantly executed (or only be executed in a limited number explicitly specified) on the devices with the same specifications.

\begin{lstlisting}[
caption={{An Example of test cases for API  \emph{NotificationChannel.\\setDescription(String)}.}},
label=code:generated_test_case,
firstnumber=1,abovecaptionskip=1pt, belowcaptionskip=3pt, aboveskip=2pt, belowskip=2pt]
@RunWith(AndroidJUnit4.class)
public class TestCase_Example {
 @Test
 public void testCase() throws Exception {
  Context var2 = InstrumentationRegistry.getTargetContext();
  String var3 = var2.getString(2131558623);
  CharSequence var1 = (CharSequence)var3;
  NotificationChannel var4 = new NotificationChannel("Reminder", var1, 4);
  String var5 = new String();
  var4.setDescription(var5);
 }
}
\end{lstlisting}

In addition, we provide a concrete test case example (as shown in Listing~\ref{code:generated_test_case}) for better comprehensibility. This test case is generated by JUnitTestGen and is specifically designed to test API \emph{NotificationChannel.setDescription(String)} at line 10. With the help of LazyCow, this test case would be allocated and dispatched to various device clusters for compatibility issues testing.

\textbf{(7) Test Case Packaging \& Dispatch.}
Once the test cases to be run on each client are assigned, \tool{} packages and dispatches the test cases to corresponding clients. \tool{} integrates hot-swap technique \cite{hotswap} to apply code changes (i.e., assigned test cases) to the client without re-installing the \tool{} app. To achieve this, we monitor file (i.e., test cases) changes and run a custom Gradle task to generate .dex files for the modified classes only. After that, another Gradle command was used to generate the newly .dex files and package them into an APK and then send the APK back to the client. \tool{} client is then able to reload these newly assigned test classes and invoke them by using reflection calls.


\textbf{(8) API Compatibility Analysis.}
After the test cases have been executed on various Android devices, \tool{} collects all the execution results from the clients and stores them in a database. \tool{} records if each test case is successfully executed on the device, and if not, the corresponding exception information or error messages (e.g., Assertion error message) are also recorded.

The API Compatibility Analysis module then evaluates the results across all devices to identify API-induced compatibility issues. We consider an Android API has a compatibility issue if its execution results are inconsistent on different Android devices. To be more specific, a given Android API is deemed to have a compatibility issue if any of the following happens: (1) One test case fails on certain devices but runs successfully on others; or (2) The test case throws different errors or exceptions on different device configurations (e.g., throws \emph{NoSuchMethodError} on some versions, while throws \emph{SecurityException} on others). Based on the comparative analysis results, \tool{} can flag vendor-specific, model-specific, and Android version-specific compatibility issues for Android APIs, which have been long-time challenges that are not yet resolved by existing approaches proposed specifically to detect compatibility issues in Android devices (such as CiD, FicFinder, etc.).

\section{Evaluation}



We  investigate the feasibility and effectiveness of detecting  compatibility issues in Android devices with \tool{} by answering the following research questions:


\begin{itemize}
\item {\bf RQ1: }{How efficient is \tool{} in supporting \\crowdsourced unit testing?}

\item {\bf RQ2: }{How effective is \tool{} in discovering \\fragmentation-induced Compatibility Issues?}

\item {\bf RQ3: }{How does \tool{} compare with existing tools in detecting compatibility issues?}

\item {\bf RQ4: }{Is \tool{} useful in practice from users' perspective? } 
\end{itemize}




\subsection{RQ1 -- How efficient is LazyCow in supporting crowdsourced
unit testing?}

\begin{table}[!h]
\centering
\caption{Device Information.}
\label{tab:device_information}
\scriptsize
\begin{tabular}{ |c | c| c| c| c|} 
\hline
ID & Brand & Model  & API Level & SoC \\
\hline
1 & \multirow{3}{*}{\textbf{Samsung}} & SM-A305YN & 30 & exynos7904 \\
2 & &  SM-A520F & 26 & exynos7880 \\
3 & &  SM-A705YN & 29 & qcom \\
\hline
4 & \multirow{3}{*}{\textbf{Huawei}} & VOG-L09 & 28 & kirin980 \\
5 &  & HMA-L29 & 29 & kirin980 \\
6 &  & JKM-AL00b & 28 & kirin710 \\
\hline
7 & \multirow{2}{*}{\textbf{Xiaomi}} & Redmi 8A & 29 & qcom \\
8 & & MI 8 UD & 29 & qcom \\
\hline
9 & \textbf{OnePlus}  & ONEPLUS A3010 & 28 & qcom \\
\hline
10 & \textbf{Honor} &  COL-AL10 & 29 & kirin970 \\
\hline
11 & \textbf{Meizu} & Meizu 17 & 30 & qcom \\
\hline
\end{tabular} 
\end{table}

\begin{table*}[!h]
\centering
\caption{Experimental results of the executed test cases on different allocation strategies.}
\label{tab:executed_results_dispatch_strategy}
\footnotesize
\begin{tabular}{| c |c | c |c| c| } 
\hline
\multirow{2}{*}{\textbf{Batch Size}} &
 \multicolumn{2}{c}{\textbf{Discard Strategy}} &
  \multicolumn{2}{|c|}{\textbf{Rebuild Strategy}}  \\
  \cline{2-5} 
  & \textbf{Avg. Executed Cases (coverage)} & \textbf{Avg. Execution Time} & \textbf{Avg. Executed Cases (coverage)} & \textbf{Avg. Execution Time}  \\
\hline
 100 & 5,008 (92.7\%) & 4,147\textit{s} & 5,339 (98.9\%) & 9,045\textit{s}\\
\hline
 500 & 4,884 (90.4\%) & 1,081\textit{s} & 5,183 (96.0\%) & 2,110\textit{s}\\
\hline
1,000 & 4,479 (82.9\%) & 825\textit{s} & 5,091 (94.3\%)& 1,499\textit{s}\\
\hline
\end{tabular} 
\end{table*}

\subsubsection{Experimental settings} 
Keeping in mind that the \tool{} runs the test cases when the devices are idle, hence, instead of loading all the test cases to the devices and executing them at one time, we designed different allocation strategies to build and dispatch test cases to the devices in batches. The allocation strategy specifies the following two factors that may affect the efficiency of \tool{}: the size of each batch (i.e., the number of test cases built and dispatched every time) and the crash handle approach (i.e., how the remaining test cases handled when a crash occurs in a batch). 

In this experiment, we vary the batch sizes to 100, 500, and 1000 to explore the impact of batch size on the execution efficiency of \tool{}. 
For the crash handling approach, we propose two strategies when a crash happens in a batch, namely the \textit{Rebuild Strategy} and the \textit{Discard Strategy}. In the Rebuild Strategy, \tool{} rebuilds the unexecuted test cases into the next batch. In the Discard Strategy, \tool{} discards the remaining test cases in a batch when a crash occurs and will restart, when available, by executing the next batch. For example, assume a batch size of 1000 is chosen. When a crash happens in the 10th test case, the Rebuild Strategy requests the following 1000 test cases from the 10th case (i.e., from 11th to 1010th test cases) as a new batch, while the \textit{Discard Strategy} skips the 990 test cases that are unexecuted and requests a new batch containing the 1001st to 2000th test cases. Intuitively, the rebuild strategy may increase the execution rate of the test cases but introduce additional overhead in building the test cases. In contrast, the discard strategy may skip some test cases but not introduce additional building and execution costs.

We use 11 Android smartphones from various manufacturers with different Android OS versions in the experiment. These 11 devices are not selected by the authors but came from real-world users. Specifically, we advertised LazyCow online and recruited Android users to download and install LazyCow on their devices. In total, we recruited 11 participants and thus had 11 Android devices involved in this work. As part of our future work, we plan to recruit further users with different devices to detect more compatibility issues.  Table~\ref{tab:device_information} summarises the detailed device information, including brand, model number, Android API level, and SoC (system-on-a-chip).

We prepare a set of test cases to evaluate the efficiency of \tool{}. We collect test cases by two means: (1) Assembling unit tests from the Android Open Source Project (AOSP) codebase; (2) Using the tool JUnitTestGen~\cite{sun2022mining} to automatically generate test cases through mining existing API usages in real-world apps. Briefly, this tool applies inter-procedural data-flow analysis to identify the API usage, including API caller instance inference and API parameter value inference.
In terms of test case generation, we randomly select 1,000 apps from AndroZoo~\cite{liu2020androzooopen} for each target SDK version (as specified in the manifest) between 21 (i.e., Android 5.0) and 29 (i.e., Android 10.0\footnote{The latest version at the time when we conducted this study.}). Here, we select 1,000 apps for each target SDK version because compatibility issues mainly lie in the evolution of APIs on different Android SDK versions~\cite{li2018cid}. In addition, JUnitTestGen performs static program analysis to automatically generate test cases from existing API usages in real-world apps. Although the test cases are generated based on API usage, the output of JUnitTestGen are minimal executable code snippets, which are also regarded as test cases. {Following the step of test cases collection, we notice that for a given target API, multiple test cases can be generated based on its various usages in real-world apps. However, it is time-consuming to execute all test cases that share the same target API. Thus, it is necessary to filter them out to save subsequent testing time and resources. Here, we first obtain the API invocation sequence for each test case and then select the smallest-scale one, which has the least number of API invocation sequences. Specifically, a Java-written script has been applied to this process, enabling an automatic approach without any human intervention and and with limited time consumption. In total, we successfully collected 5,401 test cases (covering 5,401 unique Android APIs), including 1,203 from AOSP codebase and 4,198 generated by JUnitTestGen.}


To investigate \tool{}'s efficiency of dispatching and executing unit tests, we install \tool{} on all the devices and then test it across different allocation strategies. For each run, we record the number of successfully executed test cases and the execution time for evaluation. Note that we consider a test case is successfully executed if it is successfully invoked, regardless of the execution results (e.g., exception or crash).

\begin{table}[!h]
\centering
\caption{Experimental results of the executed test cases for each device under batch size 100.}
\label{tab:executed_results_each_device_100}
\footnotesize
\begin{tabular}{ | c | c | c | c | c |} 
\hline
\multirow{2}{*}{\textbf{Device ID}} &
\multicolumn{1}{|c|}{\textbf{\# Executed}} & \textbf{\# Execution} & \textbf{Executed} & \textbf{Execution} \\
& \multicolumn{1}{|c|}{\textbf{Cases (D)}} & \textbf{Time (D)} & \textbf{Cases (R)} & \textbf{Time (R)} \\
\hline
1 & 4,629 & 5,783\textit{s} & 5,247  & 7,681\textit{s}\\
\hline
2 & 5,363 & 2,717\textit{s} & 5,368  & 7,043\textit{s}\\
\hline
3 & 4,524 & 7,284\textit{s} & 5,374 & 10,687\textit{s}\\
\hline
4 & 4,894 & 2,981\textit{s} & 5,364 & 5,957\textit{s}\\
\hline
5 & 5,174 & 5,754\textit{s} & 5,374 & 10,353\textit{s}\\
\hline
6 & 5,023 & 2,931\textit{s} & 5,180 & 8,887\textit{s}\\
\hline
7 & 4,395 & 3,232\textit{s} & 5,374 & 10,983\textit{s}\\
\hline
8 & 5,373 & 3,802\textit{s} & 5,374 & 8,999\textit{s}\\
\hline
9 & 5,147 &  2,534\textit{s} & 5,360 & 5,898\textit{s}\\
\hline
10 & 5,374 & 6,456\textit{s} & 5,374 & 14,018\textit{s}\\
\hline
11 & 5,191 & 2,147\textit{s} & 5,342 & 8,988\textit{s}\\
\hline
\end{tabular} 
\end{table}

\begin{table}[!h]
\centering
\caption{Experimental results of the executed test cases for each device under batch size 500.}
\label{tab:executed_results_each_device_500}
\footnotesize
\begin{tabular}{ | c |c | c |c| c|} 
\hline
\multirow{2}{*}{\textbf{Device ID}} &
\multicolumn{1}{|c|}{\textbf{\# Executed}} & \textbf{\# Execution} & \textbf{Executed} & \textbf{Execution} \\
& \multicolumn{1}{|c|}{\textbf{Cases (D)}} & \textbf{Time (D)} & \textbf{Cases (R)} & \textbf{Time (R)} \\
\hline
1 & 4,831 & 1,544\textit{s} & 5,346  & 1,326\textit{s}\\
\hline
2 & 5,042 & 652\textit{s} & 5,042  & 978\textit{s}\\
\hline
3 & 4,929 & 1,375\textit{s} & 5,373 & 1,794\textit{s}\\
\hline
4 & 4,518 & 626\textit{s} & 4,928 & 2,702\textit{s}\\
\hline
5 & 5,374 & 1,216\textit{s} & 5,097 & 3,041\textit{s}\\
\hline
6 & 5,080 & 574\textit{s} & 5,032 & 1,795\textit{s}\\
\hline
7 & 4,873 & 2,340\textit{s} & 5,374 & 1,701\textit{s}\\
\hline
8 & 5,374 & 691\textit{s} & 5,374 & 2,740\textit{s}\\
\hline
9 & 4,267 &  641\textit{s} & 4,735 & 1,123\textit{s}\\
\hline
10 & 4,929 & 1,646\textit{s} & 5,374 & 2,714\textit{s}\\
\hline
11 & 4,507 & 584\textit{s} & 5,343 & 3,301\textit{s}\\
\hline
\end{tabular} 
\end{table}

\begin{table}[!h]
\centering
\caption{Experimental results of the executed test cases for each device under batch size 1000.}
\label{tab:executed_results_each_device_1000}
\footnotesize
\begin{tabular}{ | c |c | c |c| c|} 
\hline
\multirow{2}{*}{\textbf{Device ID}} &
\multicolumn{1}{|c|}{\textbf{\# Executed}} & \textbf{\# Execution} & \textbf{Executed} & \textbf{Execution} \\
& \multicolumn{1}{|c|}{\textbf{Cases (D)}} & \textbf{Time (D)} & \textbf{Cases (R)} & \textbf{Time (R)} \\
\hline
1 & 3,802 & 992\textit{s} & 5,346  & 970\textit{s}\\
\hline
2 & 5,041 & 895\textit{s} & 5,043  & 983\textit{s}\\
\hline
3 & 4,392 & 851\textit{s} & 5,374 & 974\textit{s}\\
\hline
4 & 4,031 & 698\textit{s} & 4,598 & 1,303\textit{s}\\
\hline
5 & 5,374 & 755\textit{s} & 5,374 & 1,314\textit{s}\\
\hline
6 & 5,050 & 847\textit{s} & 5,139 & 979\textit{s}\\
\hline
7 & 4,334 & 1,115\textit{s} & 5,374 & 1,792\textit{s}\\
\hline
8 & 5,374 & 566\textit{s} & 5,374 & 2,765\textit{s}\\
\hline
9 & 3,715 &  965\textit{s} & 3,659 & 1,307\textit{s}\\
\hline
10 & 5,374 & 859\textit{s} & 5,374& 1,240\textit{s}\\
\hline
11 & 2,787 & 533\textit{s} & 5,345 & 2,864\textit{s}\\
\hline
\end{tabular} 
\end{table}

\subsubsection{Results} Table~\ref{tab:executed_results_dispatch_strategy} demonstrates the coverage and time overhead of executing test cases under different allocation strategies. When \textbf{increases the batch size from 100 to 500, the number of successfully executed test cases on each device decreases slightly}, from 5,008 to 4,884 (with coverage decreases from 92.7\% to 90.4\%) on the \textit{Discard strategy}, and from 5,339 to 51,83 (with coverage decreases from 98.9\% to 96.0\%) on the \textit{Rebuild strategy}. However, \textbf{the time overhead reduces dramatically} from 4,147 seconds (i.e., 0.83 seconds per case) to 1,081 seconds (i.e., 0.22 seconds per case) on the \textit{Discard strategy} and 9,045 seconds (i.e., 1.69 seconds per case) to 2,110 seconds (i.e., 0.41 seconds per case) on the \textit{Rebuild strategy}. When further increasing the batch size from 500 to 1,000, the time overhead reduces not as significant as when the batch size increases from 100 to 500 on both strategies (i.e., from 0.22 to 0.18 and 0.41 to 0.29 seconds per case, respectively), with a slightly decreased coverage (i.e., by 7.5\% and 1.7\%, respectively). 

The \textbf{\textit{Rebuild} strategy has slightly higher coverage than the \textit{Discard} strategy}, (96.0\% compared with 90.4\% when batch size is 500). However, the \textbf{\textit{Rebuild} strategy doubles the execution time} (0.41 seconds compared with 0.22 seconds per case). This is mainly caused by the frequent rebuilding of the test cases when crashes occur. 
These results suggest that \tool{} can achieve high coverage in various settings in a reasonable time. However, users have to make a trade-off between getting a higher coverage and reducing the time overhead. For example, the \textit{Discard Strategy} is significantly more efficient compared with the \textit{Rebuild Strategy}. A larger batch size is preferred if the device has a longer idle window.

\vspace{5pt}

In addition, we further break down the experimental results of executed test cases/execution time for each device under different batch size and allocation strategies in Table~\ref{tab:executed_results_each_device_100},~\ref{tab:executed_results_each_device_500} and ~\ref{tab:executed_results_each_device_1000}, where ``D'' represents the \textit{Discard Strategy} while ``R'' represents the \textit{Rebuild Strategy}. For each batch size, the overall results are consistent with what we find in Table~\ref{tab:executed_results_dispatch_strategy}. More specifically, the number of executed test cases under \textit{Rebuild Strategy} is larger than that with \textit{Discard Strategy} for each device, while the time overhead is the contrary (i.e., the \textit{Rebuild Strategy} takes more time to execute tests compared to the \textit{Discard Strategy}). Moreover, the increase of batch size helps reduce the number of the executed test cases and the execution time enormously.

\begin{figure}[!h]
    \centering
    \setlength{\belowcaptionskip}{3pt}
    \setlength{\abovecaptionskip}{3pt}
    \includegraphics[width=\linewidth]{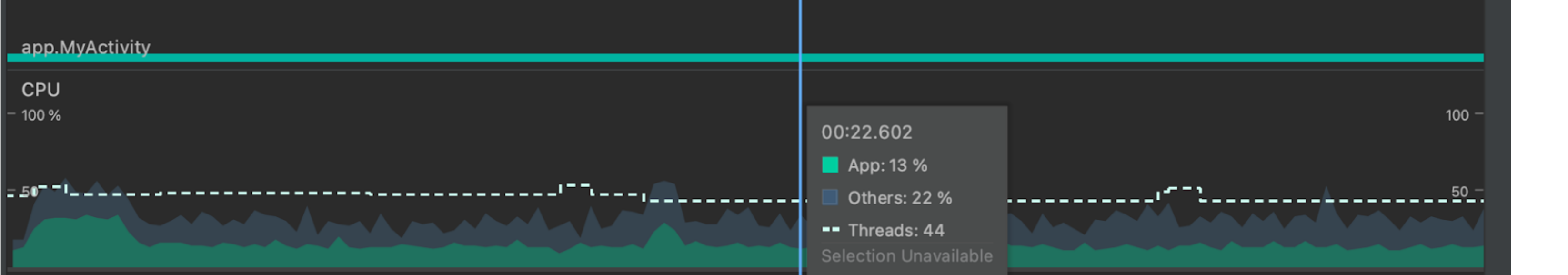}
	\caption{The CPU usage of Device 2.}
    \label{fig:CPU}
\end{figure}

\vspace{5pt}

\begin{figure}[!h]
    \centering
    \setlength{\abovecaptionskip}{3pt}
    \includegraphics[width=\linewidth]{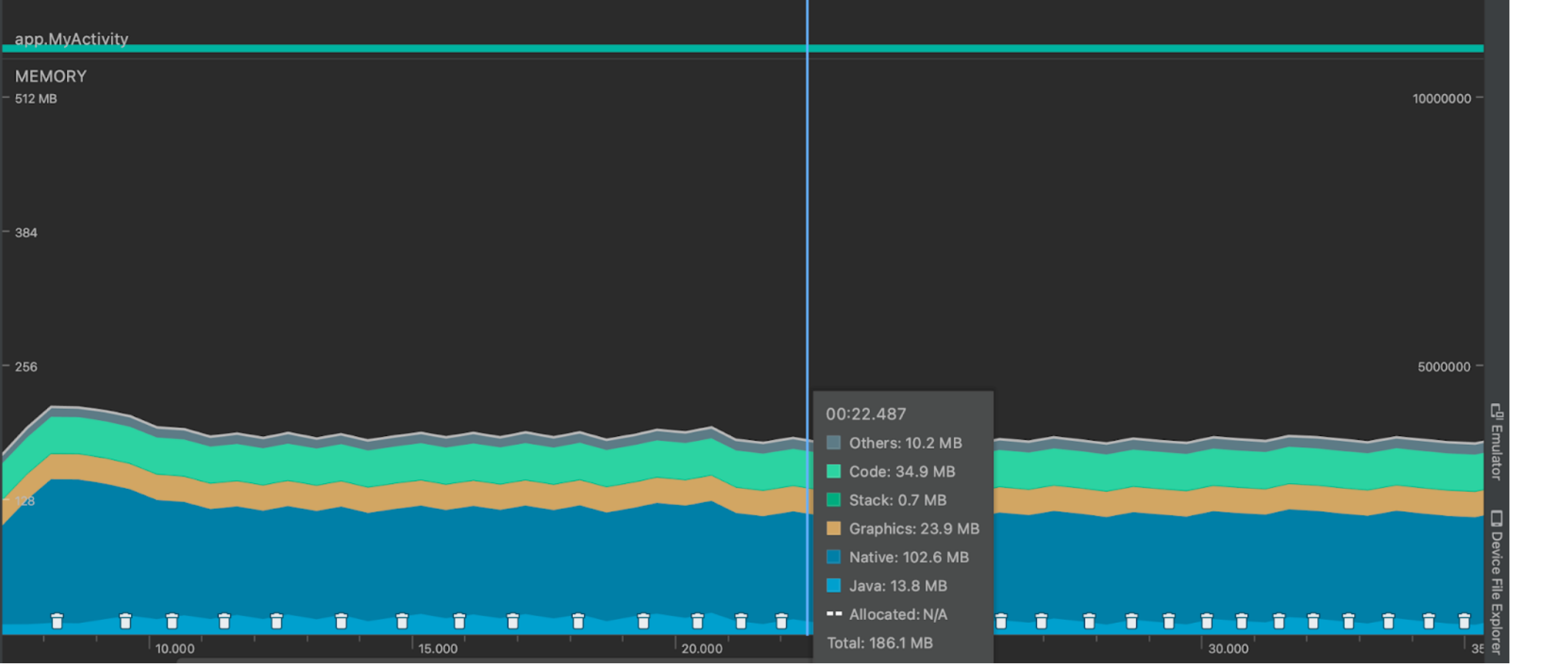}
	\caption{The memory usage of Device 2.}
    \label{fig:memory}
\end{figure}

\begin{figure}[!h]
    \centering
    \setlength{\abovecaptionskip}{0pt}
    \includegraphics[width=\linewidth]{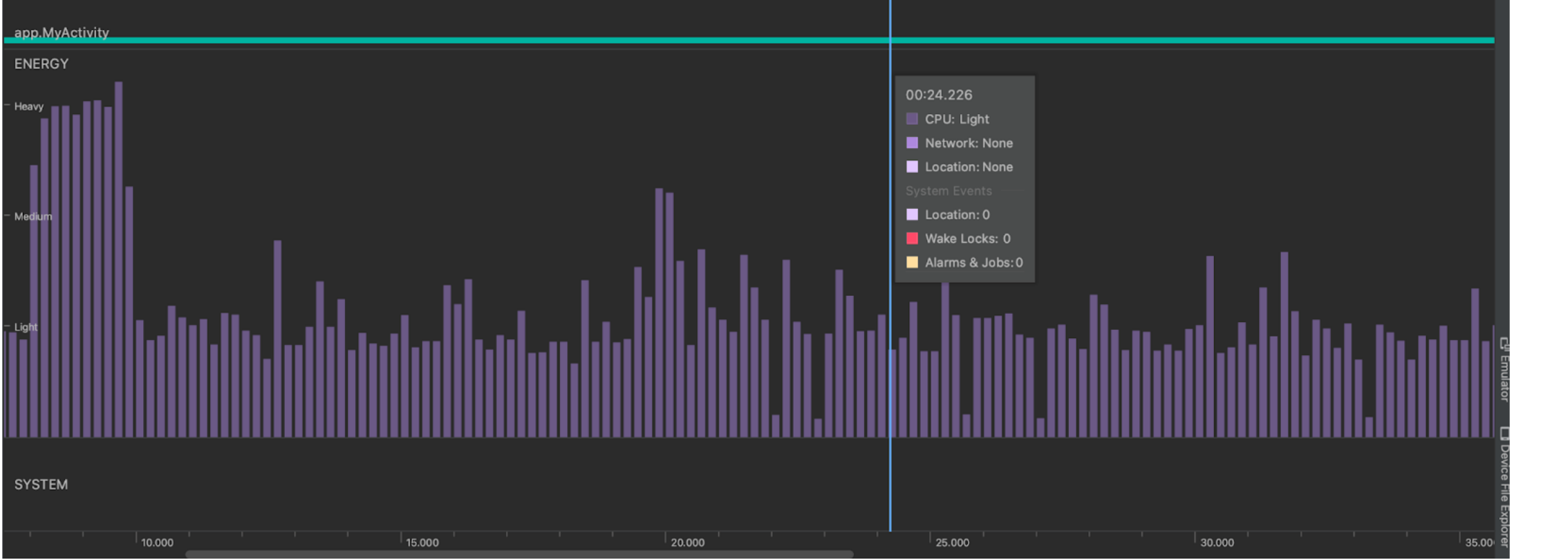}
	\caption{The energy consumption of Device 2.}
    \label{fig:energy}
\end{figure}

We further investigate the efficiency of \tool{} by examining its CPU/memory/Energy usage when running \tool{} on top of Android devices. Specifically, we picked up a random device (i.e., Device 2 with Samsung, API level 26) and presented the CPU/memory/Energy usage when running \tool{} on top of it. The following three figures~\ref{fig:CPU},~\ref{fig:memory},~\ref{fig:energy} summarise the CPU usage, memory allocations, and energy consumption, indicating that \tool{} is a resource-friendly app, which has a reasonable amount of energy consumption. Specifically, the occupation of CPU consumption is quite stable, around 13\% when running \tool{} as shown in Figure~\ref{fig:CPU}. In addition, the memory usage is presented in figure~\ref{fig:memory}, from which the 'code' category shows the Memory that \tool{} uses for code and resources, such as dex bytecode, optimised or compiled dex code, .so libraries, and fonts. As we can see, the memory usage of \tool{} is 34.9M, which is a low-memory consumption. Moreover, as indicated in figure~\ref{fig:energy}, the default view for the energy profiler shows the energy consumption of \tool{}, which is quite light and stable. 

\begin{tcolorbox}[title=\textbf{Answers to RQ1}, left=2pt, right=2pt,top=2pt,bottom=2pt]
\tool{} is efficient in supporting crowdsourced unit tests. It achieves high coverage under various settings (e.g., number of test cases dispatched every time and whether to rebuild or discard the remaining test cases when a crash happens). The users can further make a trade-off between getting a higher coverage and reducing the time overhead to suit their needs.
\end{tcolorbox}

\subsection{RQ2 -- How effective is \tool{} in discovering fragmentation-induced Compatibility Issues? }


Our second research question concerns the effectiveness of \tool{} in detecting compatibility issues in Android devices. 
Based on the definition of incompatible issues in work proposed by Cai et al.~\cite{cai2019large}, We consider that an Android API will introduce compatibility issues if its execution results across different device configurations (e.g., model, OS version, etc.) are inconsistent. To be more specific, an Android API is regarded as containing a compatibility issue if any of the following happens: (1) One test case fails on specific device configurations but runs successfully on others; or (2) The test case throws different errors or exceptions on different device configurations (e.g., throws \emph{NoSuchMethodError} on some configurations, while throws \emph{SecurityException} on others).

\subsubsection{Experiment settings} 
Except for the 11 real-world devices in RQ1, we also include Android emulators with API level 26/28/29/30 in this research question. The reason why we select these specific four API levels is because they should be consistent with the versions appearing in the real-world devices. The test cases involved are the same as RQ1.

\subsubsection{Results}
After analyzing and comparing the test case execution results based on the aforementioned three rules, \textbf{\tool{} detects 393 Android APIs that may have compatibility issues}. To confirm whether the APIs identified by \tool{} indeed have compatibility issues, we manually examine the APIs' implementation in Android framework source code on different Android SDK versions. In this step, we manually examined the 393 APIs, all of which are confirmed to be true positives, suggesting a 100\% true positive rate. It is also worth mentioning that the high true positive rate benefits from the dynamic testing technique, which provides actual evidence to pinpoint compatibility issues. This result shows that \tool{} is indeed capable of identifying compatibility issues on real-world Android devices. 
Unfortunately, due to the lack of ground truth, we cannot evaluate if there are false-negative results (i.e., compatibility issues that are missed by \tool{}). 

We summarize the compatibility issues into two major types based on their specific errors/exceptions. We manually investigate the exception type and its corresponding API signature and implementation in AOSP to examine whether it is caused by API signature change or API implementation change. The definitions of Signature-based compatibility issues and Semantics-based compatibility issues are based on our observation and are detailed as follows.
\begin{itemize}
\item \textbf{Type 1: Signature-based compatibility issues.} This type refers to the incompatibility caused by API deprecation or the change of API signature, such as introducing new APIs to the SDK, changing existing APIs' parameters or their return types, etc. For example, \emph{NoClassDefFoundError}, \emph{NoSuchMethodError} and \emph{NoSuchFieldError} are typical signature-based exceptions.

\item \textbf{Type 2: Semantics-based compatibility issues.} This type implies that the API has consistent signatures over different android SDK versions, but its implementation has been changed. For example, \emph{RuntimeException},\emph{SecurityException} and \emph{NullPointerException} are typical semantics-based exceptions.
\end{itemize}

\begin{figure}[!h]
    \centering
    \includegraphics[width=\linewidth]{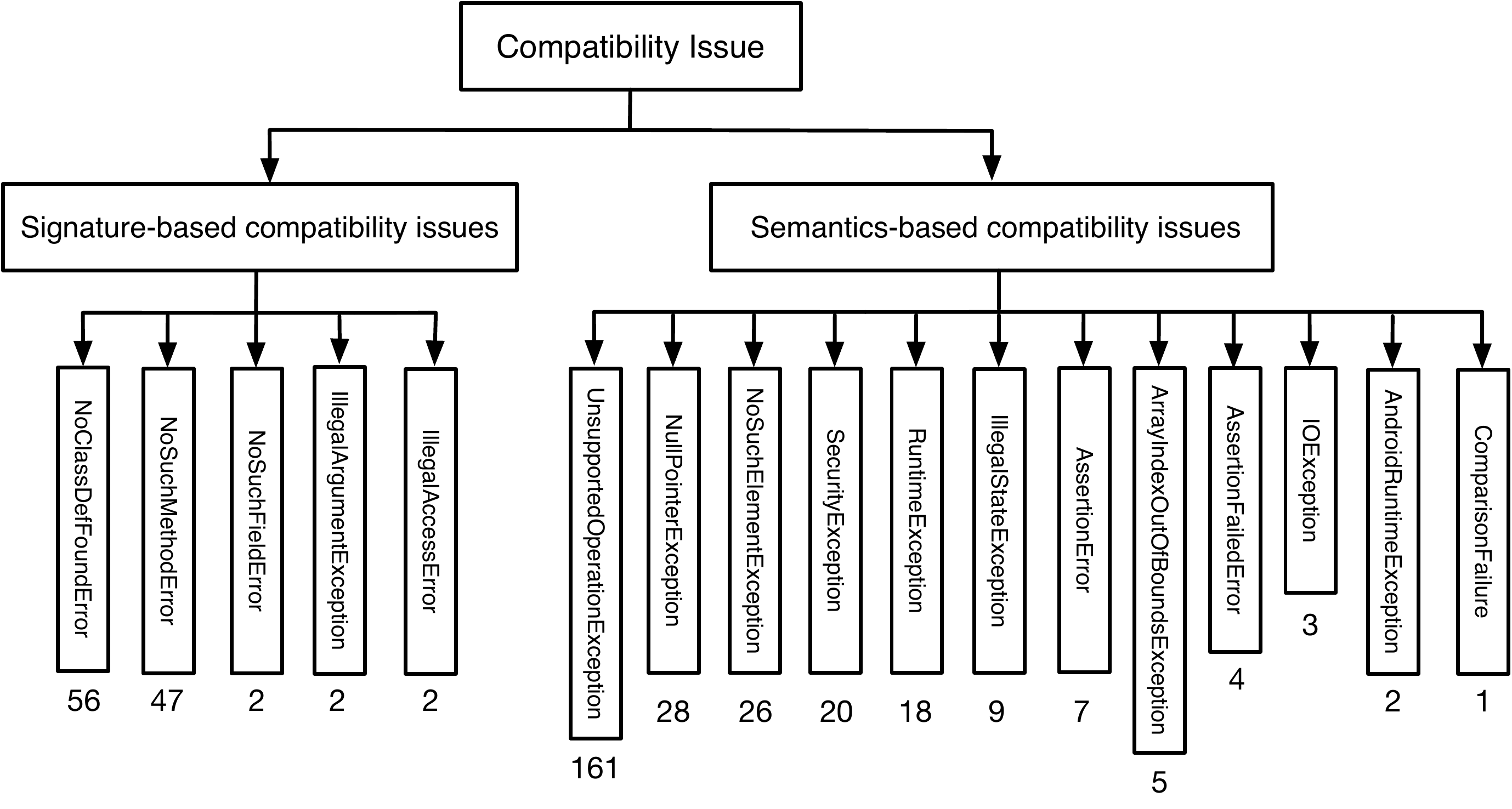}
	\caption{The category of error/exception types associated with compatibility issues}
    \label{fig:ExceptionType}
\end{figure}

Figure~\ref{fig:ExceptionType} summarizes all the possible errors/exceptions caused by both types in the dataset. Among the 393 identified compatibility issues, \textbf{109 of them belong to signature-based issues and 284 are semantic-based issues}. Among signature-based issues, \emph{NoClassDefFoundError}, \emph{NoSuchMethodError} and \emph{NoSuchFieldError} are the most frequently occurred errors.  \emph{UnsupportedOperationException} and \emph{SecurityException} appear the most frequently in terms of semantic-based issues. Table~\ref{tab:Compatibility_each_device} further breaks down the number of compatibility issues for each device. For example, \tool{} identified 193 signature-based compatibility issues and 150 semantic-based compatibility issues on the Samsung device with Android API level 30 (i.e., Device 1). In addition, the number of detected compatibility issues is quite stable on real-world devices (i.e., Device 1 to 11), with 338 on average for each device. Despite this, we find out that the number of compatibility issues on Android emulators increases as the API level evolves consecutively. For instance, the number of compatibility issues increases from 192 to 308 on API level 26 to 28. As revealed by Li et al.~\cite{li2018cid}, the fast evolution of the Android framework has
indeed deprecated/removed/introduced a lot of APIs that will likely induce compatibility issues.

\begin{table}[!h]
\centering
\caption{The number of compatibility issues identified on each device.}
\label{tab:Compatibility_each_device}
\footnotesize
\scalebox{0.9}{
\begin{tabular}{ | c |c | c |c| c| } 
\hline
\multicolumn{1}{|c}{\textbf{Device}} &
\multicolumn{1}{|c}{\textbf{API}} &
\multicolumn{1}{|c|}{\textbf{\# Signature-based}} & \textbf{\# Semantics-based}  &
\multirow{2}{*}{\textbf{Count}} \\
 \multicolumn{1}{|c|}{\textbf{ID}} &  \multicolumn{1}{c|}{\textbf{Level}} 
& \multicolumn{1}{c|}{\textbf{incompatibility}} &  \multicolumn{1}{c|}{\textbf{incompatibility}} & \\
\hline
1 & 30 & 193 & 150 & 343 \\
\hline
2 & 26 & 228 & 118 & 346 \\
\hline
3 & 29 & 243 & 104 &  347 \\
\hline
4 & 28 & 241 & 85 &  326 \\
\hline
5 & 29 & 210 & 123 & 333 \\
\hline
6 & 28 & 235 & 88 & 323 \\
\hline
7 & 29 & 112 & 203 & 315\\
\hline
8 & 29 & 120 & 243 &  363\\
\hline
9 & 28 & 231 & 116 & 347 \\
\hline
10 & 29 & 252 & 83 & 335 \\
\hline
11 & 30 & 221 & 124 &  345 \\
\hline
Emulator & 26 & 124 & 68 &  192 \\
\hline
Emulator & 28 & 203 & 105 &  308 \\
\hline
Emulator & 29 & 214 & 117 &  331 \\
\hline
Emulator & 30 & 223 & 122 &  345 \\
\hline
\end{tabular}}
\end{table}

The following provides an illustrative case study for each type of compatibility issue.

\textbf{Case Study 1: Signature-based Compatibility Issue.}
The API \emph{android.security.keystore.KeyGenParameterSpec.\\Builder\#setUnlockedDeviceRequired} has been reported to contain a signature compatibility issue. The corresponding test case throws \emph{NoSuchMethodError} on the device with API level 26 but can be successfully executed on other devices whose API level is greater or equal to 28. This result indicates that it would cause exceptions if an API is invoked on a certain API level earlier than it has been introduced.

\textbf{Case Study 2: Semantic-based Compatibility Issue.}
The API \emph{android.util.LongSparseArray\#valueAt} has been reported to contain a semantics-based compatibility issue. The corresponding test case throws \emph{ArrayIndexOutOfBoundsException} on API level 29 and 30 while successfully executed on the other devices. We manually looked into its source code in the Android codebase and found that the actual implementation of this API has been changed since API level 29 but the signature of the API remains the same (i.e., the return type and the parameters). 
It is worth noting that static methods such as CiD \cite{li2018cid} cannot detect such a case.

\begin{figure}[!h]
    \centering
    \setlength{\abovecaptionskip}{0.cm}
    \includegraphics[width=\linewidth]{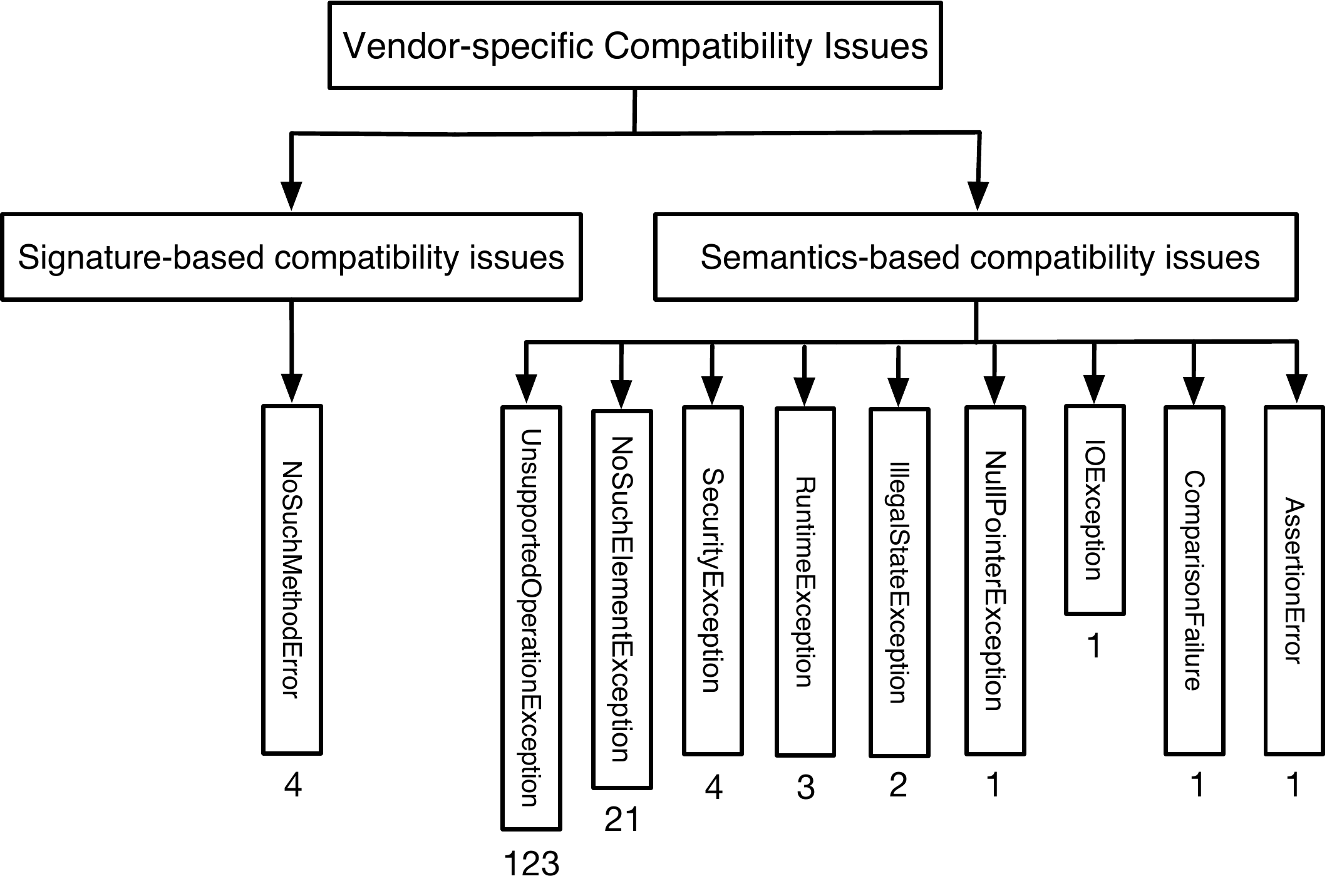}
	\caption{The category of error/exception types associated with Vendor-specific compatibility issues}
    \label{fig:vendor_specific_catgory_exception}
\end{figure}
\begin{figure}[!h]
    \centering
    \setlength{\abovecaptionskip}{0.cm}
    \includegraphics[width=0.6\linewidth]{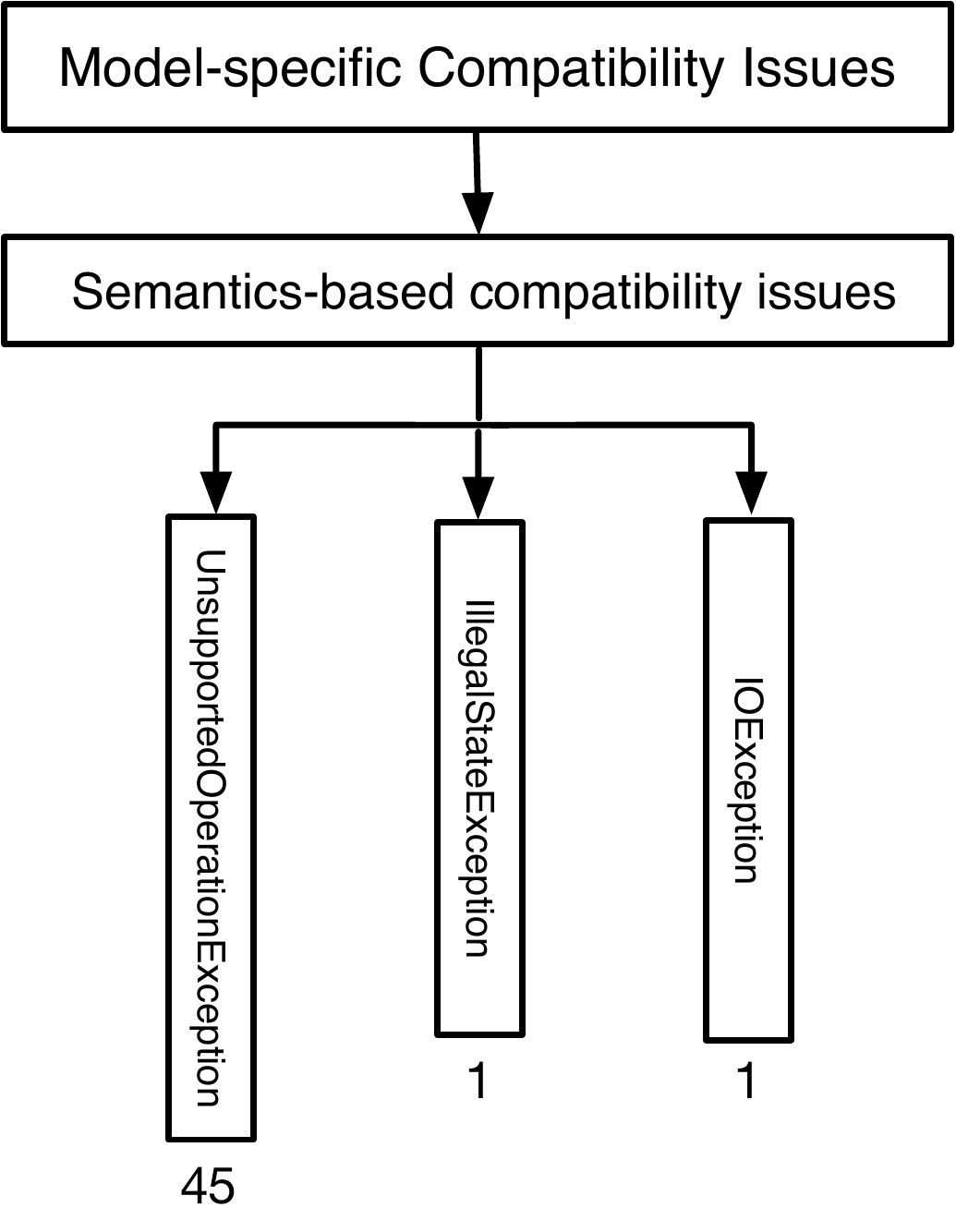}
	\caption{The category of error/exception types associated with Model-specific compatibility issues}
    \label{fig:model_specific_catgory_exception}
\end{figure}

\textbf{Vendor/Model-specific Compatibility Issues.}
We find that some compatibility issues are vendor/model-specific, which are caused by unique AOSP customizations. We consider that an Android API will introduce a vendor-specific compatibility issue if its execution results are inconsistent across different brands with the same API level, and a model-specific compatibility issue if its execution results are inconsistent across the devices from the same brand and with the same API level.
As a result, \textbf{\tool{} is able to detect 161 APIs with vendor-specific compatibility issues and 47 model-specific compatibility issues}, by comparing the execution results on different brands/models of devices with the same Android API level.
Figure~\ref{fig:vendor_specific_catgory_exception} and \ref{fig:model_specific_catgory_exception} further illustrates the possible errors/exceptions that exist in vendor- and model-specific compatibility issues, respectively. It is observed that most vendor- and model-specific compatibility issues are semantically based. For instance, most incompatible APIs throw \emph{UnsupportedOperationException} on the specific device(s), which suggests that the requested operation is not supported.  

According to Wu et al.~\cite{wu2013impact}, such vendor- and model-specific issues are significant and on the whole responsibility for the bulk of the security problems. 
For example, in our experiments, \tool{} identified two severe crashes that occur on Samsung and Huawei devices due to model-specific compatibility issues. Specifically, we observe a \emph{tgkill} native crash when invoking the \emph{MediaPlayer} on the Samsung device with Android API level 26 (Android 8.0). This problem is caused by the bugs in the native library libhwui.so in Samsung SM-A520F. This problem has attracted several online discussions, such as on the StackOverflow~\cite{stackoverflowSamsung}.
\tool{} also reveals that hidden APIs are likely to crash specifically on Huawei devices. Android Hidden APIs are classes, methods, and resources that Google hides from app developers to maintain stability. These APIs are located in the android.jar file with the @hide Javadoc attribute. Google hides them intentionally to avoid compatibility issues because the implementation of these APIs is prone to change in future versions. However, it is easy to bypass Google’s hidden API restrictions, as demonstrated on Github~\cite{Githubhiddenapi}. According to Li et al.~\cite{li2016accessing}, app developers are interested in harnessing hidden APIs before Google releases
them as public APIs in the future. The prevalence of hidden APIs may introduce crashes on Huawei devices, leading to poor user experiences. The following demonstrates two case studies for vendor-specific and model-specific compatibility issues.

\textbf{Case Study 3: Vendor-specific Compatibility Issue.}
The API \emph{android.animation.ValueAnimator\#setFrameDelay} has been reported to contain a vendor-specific compatibility issue. The corresponding test case can be successfully executed on all Huawei and Honor devices \footnote{Honor is a sub-brand of Huawei.} but failed on the devices from other vendors, including the ones with the same API levels as the devices mentioned above. We manually looked into the source code of the API in AOSP and found that the implementation of this API has been changed since API level 24, which suggests it should throw \emph{IllegalStateException} on the Android API level equal to or greater than 24. However, it can be successfully executed on the aforementioned Huawei and Honor devices, indicating that Huawei and Honor customized the implementation on this API, which may cause compatibility issues.


\textbf{Case Study 4: Model-specific Compatibility Issue.}
The API \emph{android.telephony.TelephonyManager\#getImei} has been reported to contain a model-specific compatibility issue. The corresponding test case can be successfully executed on device ID 4 (Huawei P30 Pro with Android API level 28) while throwing \emph{SecurityException} on device ID 6 (Huawei Enjoy 9 Plus with Android API level 28). By manually checking the official Android API documentation, we found that from API level 28, it is compulsory for apps to have the \emph{READ\_PHONE\_STATE} permission to access the IMEI of the device, and will throw \emph{SecurityException} if the permission is not granted. However, on Huawei device ID 4 (Huawei P30 Pro), it does not throw such an exception even when the \emph{READ\_PHONE\_STATE} permission is not granted. The inconsistent results on two devices from the same vendor and with the same Android API level suggest a model-specific compatibility issue.

\begin{tcolorbox}[title=\textbf{Answers to RQ2}, left=2pt, right=2pt,top=2pt,bottom=2pt]
Our approach is effective in automatically pinpointing and confirming API-induced compatibility issues. It also goes beyond the state-of-the-art to detect not only signature-based but also semantic-based compatibility issues. Our results also demonstrate that \tool{} can identify vendor- and model-specific compatibility issues, which may bring severe security problems to the mobile ecosystem.
\end{tcolorbox}

\subsection{RQ3 -- How does \tool{} compare with existing tools in detecting compatibility issues?}
Given that the main purpose of our work is detecting compatibility issues, both static/dynamic compatibility issues detection tools, such as CiD~\cite{li2018cid}, JUnitTestGen~\cite{sun2022mining} and a commercial-grade Compatibility Test Suite (Google CTS~\cite{CTS}), are selected as the baselines to evaluate our approach. We evaluate the performance of \tool{}, CiD, JUnitTestGen and CTS in detecting compatibility issues. Here we break down the comparative results as follows:

\textbf{Comparison with CiD.}
To compare \tool{} with a static compatibility issues detection tool, we selected the state-of-the-art tool,CiD,
which mines the historical versions of Android framework source code to find incompatible API usages, as our baseline. We ran CiD on the same dataset as used in RQ1, i.e., 10,000 apps. To make the comparison fair, we only include CiD results that have compatibility issues on the API levels contained in our experiment devices, i.e., API levels 26, 28, 29, and 30. Compared with 393 compatibility issues detected by \tool{}, CiD has identified 215 compatibility issues. Specifically, CiD failed to detect 336 issues identified by \tool{}, with 100 of which are signature-based issues, and 236 are semantic-based issues. We further analyze the results and find that the missed signature-based issues are caused by system customization that CiD cannot spot. Missing semantic-based issues is also in line with the intuition that because of lacking semantic analysis, CiD is not capable of pinpointing the compatibility issues caused by the semantic change of an API. On the other hand, \tool{} missed 158 compatibility issues detected by CiD. Of these 134 are caused by failed generated test cases for the APIs (e.g., UI-related APIs). It is non-trivial to automatically generate unit tests in general as it requires sophisticated semantic analysis to model the sequences of statement invocations. For example, UI-related APIs may involve the initialization of UI resources that cannot be done programmatically. We, therefore, argue that such false-negative results can be eliminated if valid test cases for these APIs are provided. In summary, our experimental results indicate that dynamic techniques (e.g., crowdsourced testing) can indeed be helpful to supplement the capabilities of static analysis.

\textbf{Comparison with JUnitTestGen.}
Apart from static analysis-based techniques,
a few dynamic approaches have been proposed to detect compatibility issues. To the best of our knowledge, JUnitTestGen is the one and only dynamic compatibility issues detecting tool~\cite{sun2022mining}. It mines Android API usages to generate unit test cases for pinpointing compatibility issues caused by the fast evolution of the Android framework. To this end, we evaluate the performance of \tool{} and JUnitTestGen on the same dataset in RQ1, which in total contains 10,000 Android apps. Here, to clarify, the key difference of \tool{} and JUnitTestGen is that \tool{} provide a crowdsourced testing platform while JUnitTestGen can only execute tests on Android emulators. To make the comparison fair, we present the experimental results on the API levels contained in our experiment devices, i.e., API levels 26, 28, 29, and 30.

As highlighted in table~\ref{tab:Compatibility_each_device}, \tool{} outperforms JUnitTestGen by detecting more compatibility issues on API level 26, 28, 29, and 30. Take API level 26 as an example, \tool{} successfully detects 228 signature-based compatibility issues and 118 semantics-based compatibility issues on device 2 while JUnitTestGen only pinpoints 124 signature-based compatibility issues and 68 semantics-based compatibility issues on emulator 26. The reason behind this is quite straightforward as JUnitTestGen only runs tests on emulators that deploy the original Android OS, overlooking many compatibility issues caused by vendor/model customization. Overall, the comparison results reveal the necessity of executing test cases on real-world devices for systematically finding Android fragmentation issues. Without the mechanism of executing tests on real-world devices, it is very hard to detect compatibility issues
caused by vendor/model customization. It also demonstrates that
our \tool{} approach can indeed find more diverse compatibility issues and
hence is promising to complement existing the dynamic approach.


\begin{table}[t!]
\centering
\caption{The comparison results between {\tool{}} and CTS on device 2 (i.e., Samsung with API level 26).}
\label{tab:comparison_CTS}
\resizebox{0.8\linewidth}{!}{
\begin{tabular}{|r | c |c |} 
\hline
Tool &  \tool{} &  CTS  \\
\hline
\# Test Cases & 5,401  & 3,602 \\
\hline
\# Compatibility Issues & 393  & 6  \\
\hline
\end{tabular} }
\end{table}


\textbf{Comparison with Google CTS.}
Google provides a commercial-grade test suite, CTS, which is designed for continuous compatibility issues in Android OS customization. The purpose of CTS is to ensure that the software remains compatible throughout the development process. We thus select CTS for effectiveness comparison as well. Given that CTS can only be tested in a lab environment and it is time-consuming to execute the full package of the test suite across all participants' devices, we thus evaluate \tool{} and CTS on a randomly selected device and compare their performance in compatibility issues detection. 

In total, \tool{} successfully executed 5,401 tests on the device under the discard strategy with batch size 100, while CTS only executed 3,602 tests on the actual device. We then compare the results of these executed tests to determine compatibility issues. As shown in Table~\ref{tab:comparison_CTS}, CTS only detects 6 compatibility issues, while \tool{} is capable of detecting 393 compatibility issues (covers the 6 compatibility issues discovered by CTS). We further manually check the test cases specified in CTS  and observe that the false negatives (compared with \tool{}) are mainly caused by the low coverage of Android system APIs. The test cases of CTS only cover a small amount of frequently used APIs, overlooking a bunch of uncommon APIs. On the other hand, occasionally while running the tests, a system dialogue may pop up informing the
user that the device is not responding. This alert dialogue obstructs the execution of tests, which causes them to fail. Besides, the lack of sufficient testing context (e.g., various parameter values) is another reason why CTS fails in detecting compatibility issues, especially those caused by vendor/model customization. Specifically, vendor/model customization may be triggered under narrow circumstances with specific inputs, which makes CTS insufficient by overlooking the variety of API parameter values. In contrast, \tool{} is able to detect such compatibility issues because we rely on JUnitTestGen in mining existing Android API usages to generate API-focused test cases, which retain the execution context in real-world applications.

\subsection{RQ4 -- Is \tool{} useful in practice from users' perspective?}
Apart from the efficiency and effectiveness analysis of our approach, it is also essential to investigate the users' satisfaction with LazyCow in practice.
Considering \tool{} is designed to run tests on users' devices, it is necessary to evaluate the human-perceived quality and usefulness of \tool{}.
To this end, in this research question, we focus on investigating the usefulness of \tool{} from users' perspective, which is also part of our initial attempts towards demonstrating the benefits of \tool{} based on what meets customers' fulfilment. Here, we use a novel commercial model, which uses people's Android smartphone to run \tool{}, and provides an incentive as payback (such as replacing the advertisements in the free apps). Specifically, we would collaborate with commercial app development companies on offering premium accounts (i.e., with no advertisements or exclusive content benefits) to customers if they choose to download LazyCow on their phone. In this way, from app development companies' perspective, they can get valuable compatibility issues reports and from customers' perspective, they have an incentive to download and install LazyCow.

In this work, we publicize our tool and randomly recruit real Android users to download and install LazyCow on their devices. Then, users are requested to use their mobile phone the same as their daily habit for a day. During this time, LazyCow works as a background service, requesting tests and then executing them when the device is in an idle state.
\subsubsection{Experiment settings.} To answer this RQ, we perform a customer satisfaction evaluation by leveraging Likert scale~\cite{Likertscale} metrics. It is a 5-point scale that ranges from a weaker endorsement (meaningless agreement with or approval) of the item (i.e., strongly disagree) to a stronger endorsement of the item (i.e., strongly agree). We used a Likert-scale closed question survey to ask users questions that gauge their satisfaction levels about \tool{}. Table~\ref{tab:survey} summarizes the questions we request the respondents to answer after using \tool{}, on a scale from 1-5, where 1 = strongly disagree, 2 = disagree, 3 = neutral, 4 = agree, 5 = strongly agree. 


\begin{table*}[!h]
\centering
\caption{The survey.}
\label{tab:survey}
\small
\begin{tabular}{ |c | l | c|} 
\hline
Score & Questions  & Scale\\
\hline
 \multirow{2}{*}{\textbf{CES}} & 1.It is easy to use LazyCow. & 1-5\\
 & 2.LazyCow won't affect my interaction with mobile phone. & 1-5\\
\hline
 \multirow{3}{*}{\textbf{CSS}} & 3.I am willing to download and install LazyCow to replace advertisements. & 1-5\\
 & 4.I feel happy with my experience with the LazyCow App. & 1-5\\
 & 5.I did not encounter compatibility problems after using LazyCow. & 1-5\\
\hline
\textbf{NPS} & 6.I will recommend LazyCow to family, friends, or colleagues. & 1-5\\
 \hline
\end{tabular} 
\end{table*}

The design of the survey is organized based on the following principles:
\begin{itemize}
\item \textbf{CES (Customer Effort Score):} CES is found by asking users to rate their effort levels. We obtain this score by asking users to rate the simplicity of using \tool{} on a scale from 1-5 (i.e., question 1 and 2). CES is calculated as:

\begin{scriptsize}
\begin{equation} 
\label{eqCES}
\begin{split}
CES & = \% Easy - \% Difficult
\end{split}
\end{equation}
\end{scriptsize}

\item \textbf{CSS (Customer Satisfaction Score):} We obtain this score by asking users' satisfaction when using \tool{} (i.e., question 3, 4 and 5). CSS is calculated as:

\begin{scriptsize}
\begin{equation} 
\label{eqCSS}
\begin{split}
CSS & = \frac{Number\ of\ satisfied\ customers}{Number\ of\ satisfaction \ survey \ responses} \times 100
\end{split}
\end{equation}
\end{scriptsize}

\item \textbf{NPS (Net Promoter Score):} NPS is calculated by asking users how likely they are to recommend \tool{} (i.e., question 6). NPS is calculated as:
\end{itemize}

\begin{scriptsize}
\begin{equation} 
\label{eqNPS}
\begin{split}
NPS & = \% Promoters - \% Detractors
\end{split}
\end{equation}
\end{scriptsize}

Before we conducted this experiment, we obtained ethics approval from the Research Ethics Committee of Monash University\footnote{The project ID is 30641.}.
We invited 11 Android users (i.e., randomly invited Android users who are interested in using \tool{} as a replacement for seeing advertisements) to install \tool{} and obtain feedback by answering the survey questions. The participants are recruited via online advertising. Each of them is asked to download and install \tool{} on their Android device and then use the phone as they usually do for a day before filling out the survey. \tool{} runs in the \emph{rebuild strategy} with a batch size of 100.

\subsubsection{Results} 
We calculate the aforementioned three scores to evaluate \tool{} from users' perspective, which are elaborated as follows:

1. The CES is calculated based on question 1 and 2. 
        \begin{scriptsize}
        \begin{equation} 
        \label{CESQuestion1}
        \begin{split}
        CES_{question1}=92\% - 0\%= 92\%
        \end{split}
        \end{equation}
        \end{scriptsize}
        \begin{scriptsize}
        \begin{equation} 
        \label{CESQuestion1}
        \begin{split}
        CES_{question2}=92\% - 0\%= 92\%
        \end{split}
        \end{equation}
        \end{scriptsize}
It is observed that 92\% of responses think it is easy to use \tool{} while no candidate feels it is difficult to use.

\begin{spacing}{0.95}
2. We measure the customer satisfaction score(CSS) based on questions 3, 4 and 5. The CSS is calculated on a scale of 0-100\% (i.e., equation~\ref{eqCSS}), where 100 represents the total customer satisfaction. Only the number of respondents who rated their satisfaction with scores 4 and 5 are included.
\end{spacing}
        \begin{scriptsize}
        \begin{equation} 
        \label{CssQuestion3}
        \begin{split}
        CSS_{question3}=\frac{11}{12}\times 100 = 92\%
        \end{split}
        \end{equation}
        \end{scriptsize}
        \begin{scriptsize}
        \begin{equation} 
        \label{CssQuestion4}
        \begin{split}
        CSS_{question4}=\frac{11}{12}\times 100 = 92\%
        \end{split}
        \end{equation}
        \end{scriptsize}
        \begin{scriptsize}
        \begin{equation} 
        \label{CssQuestion4}
        \begin{split}
        CSS_{question5}=\frac{12}{12}\times 100 = 100\%
        \end{split}
        \end{equation}
        \end{scriptsize}
The results show that \tool{} obtains a high user satisfaction score and none of the participants encountered any compatibility problems when using \tool{}.

\begin{spacing}{0.95}
3. The promoters have a scale from 4 to 5, indicating they are highly likely to recommend \tool{} to others. While the detractors have scaled from 1 to 3, indicating they are unsatisfied with the recommendation. The score is calculated based on equation~\ref{eqNPS}:
\end{spacing}
        \begin{scriptsize}
        \begin{equation} 
        \label{CssQuestion3}
        \begin{split}
        NPS_{question6}=92\% - 8\% = 84\%
        \end{split}
        \end{equation}
        \end{scriptsize}
 The results illustrate that most users are positive about \tool{}.



In addition, we observe the emphasis on reasons why/why not users have the desire to install \tool{} with incentives (i.e., they will not see advertisements anymore). For instance,
one participant said that they are eager to install \tool{} because they are ``able to get rid of annoying commercial advertisements, which would be placed in a location that covers up or hides any area that I have interest in viewing during interactions.''. Similarly, another participant stated, ``It is a fair deal to stop pop-up Ads with the cost of having a task running at idle time, which won't affect daily usage.''. Beyond \tool{}'s usefulness, participants also cite its satisfaction after they used \tool{}. For example, one participant said that they are ``satisfied with the experience of using \tool{} because I didn't even feel it running and it won't disturb my other interactions.'' Finally, while we observed little quantitative evidence of negative
experiences of \tool{}, we did observe one participant stating that ``I generally don't want to install such apps because I have no idea whether it will steal my personal information without my consent.'' 
We believe such concerns would be well mitigated if the lightweight crowdsourced testing approach is directly embedded and released by phone manufacturers.
Overall, the aforementioned answers about customer experiences for using \tool{} provide strong evidence that \tool{} is useful and welcome in practice.

\begin{tcolorbox}[title=\textbf{Answers to RQ3}, left=2pt, right=2pt,top=2pt,bottom=2pt]
By investigating the customer experiences through qualitative metrics, users' satisfaction provides strong evidence that \tool{} 
is useful and welcome in practice.
\end{tcolorbox}

\section{Discussion}
We now discuss the implications of our crowdsourced framework, potential directions for future works and the comparison with CTS.

\textbf{Practical Crowdsourced Testing Framework. } 
We have presented a lightweight crowdsourced testing framework for automatically dispatching and executing test cases without disturbing users, which is able to detect compatibility issues lies on specific devices. 
Previous works~\cite{li2018cid} resort to static analysis techniques to detect compatibility issues. However, static analysis tools are known to be imprecise because they lack the actual evidence for pinpointing compatibility issues. For example, static tools tend to overlook the semantics-based compatibility issues (i.e., APIs have the same signature but the actual implementations have been customized). Thus, in this work, we leverage dynamic crowdsourced techniques to detect broader categories of compatibility issues in real-world apps rather than signature-based ones.
Although \tool{} has been proven to be effective in detecting compatibility issues in real-world devices. \tool{} could also be easily adapted to automatically diagnose issues that exist on Android devices for other purposes. \tool{} allows users to provide customized test cases corresponding to certain issues and to dynamically execute them across Android devices around the world.

\textbf{Lightweight Crowdsourced Testing Go Beyond Android Fragmentation Taming.}
Our approach performs crowdsourced testing for Android devices, which are not strongly attached to fragmentation issues. We believe it could be easily adapted to analyze other security issues, e.g., to automatically detect hackable vulnerabilities by applying customized test cases. Our platform can be directly combined with fuzzing tools for discovering more vulnerabilities or defects in Android. It hence goes beyond compatibility testing and provides a more general-purpose form of Android testing, especially for security issues detection. We plan to explore these research directions in our future work.


\textbf{Comparison with CTS.}
Google CTS is expected to be used by manufacturers to test their new devices before releasing them to the public. The manufacturers need to download and install the CTS module on a computer to which the new device under testing is physically attached and launch the CTS. The CTS itself will then be responsible for running all its included test cases on the attached devices and subsequently storing the execution results.
As long as the new device successfully passes the CTS scanning (i.e., no test cases fail), the manufacturer could consider that the device does not contain compatibility issues and hence can be released to the public. However, CTS is known as low efficient~\cite{park2013fragmentation} and heavyweight in terms of detecting compatibility issues. Here, we elaborate on the details as follows:

\begin{itemize}
\item \textbf{Low-Efficiency :} CTS is a commercial-grade test suite, and it usually takes at least several hours to execute all the tests of CTS. Consequently, the development schedule of Android device manufacturers could be seriously affected if the CTS test is involved in the daily system integration or continuous regression testing~\cite{park2013fragmentation}. Indeed, the setup steps~\cite{CTS_setup} are pretty sophisticated because it required to run on a desktop machine and execute test cases on physically connected devices. This motivates us to provide an automated tool to parallelize the test suite across many devices (and not limited to a lab environment) so that device manufacturers can benefit from a such distributed testing strategy to resolve compatibility issues to adapt to an ever-quicker iteration process.

\item \textbf{Heavyweight:} As argued by Liu et al.~\cite{liu2016concurrent}, the CTS mechanism has been long blamed for being heavily weighted for daily testing. Consider the evolution of the CTS test case database as an example. During its evolution, some new test cases might be added, or existing test cases could be updated. These changes cannot be easily included for already released devices that have previously passed CTS scans. In other words, there is no guarantee that the changed test cases will not lead to compatibility issues in publicly available devices. Similarly, the current CTS module cannot easily support the requirements of developers in evaluating their compatibility-related test cases. Such requirements have even gone beyond the scope of CTS as it requires the test cases to not only pass a single device but also all the publicly available devices, including the ones released by different manufacturers and the different versions of the same manufacturer.
Indeed, the setup steps~\cite{CTS_setup} are quite sophisticated because it needs to run on a desktop machine and execute test cases directly on connected devices. It is also time-consuming to finish one run of the test because the CTS database already contains many test cases. Yet, it also requires dedicated efforts for developers to include new test cases into the CTS's workflow.

\end{itemize}

\subsection{Threats to validity}
\label{sec:limitation}
The main limitation of our work lies in the limited scale of evaluation. Due to both the limited number of devices available (for RQ1 and RQ2) and the Android users to participate (for RQ3), our evaluation results may not represent Android smartphone vendors and users worldwide. Nevertheless, we have included top Android smartphone vendors (e.g., Samsung, Xiaomi, and Huawei)~\cite{marketshare} with various Android versions (API levels 26-30) in our evaluation, and identified real-world compatibility issues in these devices. 
It can be foreseen that \tool{} will reveal more compatibility issues by including a wider range of vendors and devices (e.g., in a real-world crowdsourced testing scenario). 

Second, the test cases currently involved in our experiments may not cover all Android framework APIs. For example, some UI-related APIs may involve the initialization of UI resources, which makes it very challenging to programmatically generate tests for them. The limitation from unit test generation may further impact the completeness in detecting API-introduced compatibility issues. However, we argue that this limitation can be alleviated by integrating other approaches for a wider range of API coverage. Also, we remind the readers that this is not the main objective of this work, while our work is mainly focused on providing a workflow of the crowdsourced testing platform to enable users to execute customized tests on real-world devices. 

Third, another challenge of our work was the scarcity of literature providing standard metrics to evaluate how effective crowdsourced testing approaches are in practice. According to~\cite{gao2019successes}, most current crowdsourced testing approaches are owned by and operated in the industry, of which the overall workflows and detailed processes have not been published in the literature. This limitation hinders more in-depth comparisons between our work and other related works. Furthermore, the community lacks standard metrics for evaluating the effectiveness of crowdsourced testing approaches, which needs further research.

Last but not least, the capability of our approach is limited by specific types of compatibility issues, leading to false negatives. According to Mahmud~\cite{mahmud2022android}, apart from compatibility issues raised by API signature/semantic changes, there exist other types of compatibility issues, such as those introduced by field evolution, callback method changes~\cite{mahmud2021android}, etc. Nevertheless, as summarised by Liu et al.~\cite{liu2022automatically}, the number of such compatibility issues is quite limited, suggesting that the impact of such cases on our approach may not be significant.

\section{RELATED WORK}
\textbf{Android fragmentation and compatibility issues.}
Numerous works~\cite{joorabchi2013real, choudhary2015automated, han2012understanding, wei2018understanding,cotroneo2020comprehensive, li2018elegant,huang2021characterizing} have revealed that Android fragmentation is an essential challenge in the Android ecosystem. Han et al.\cite{han2012understanding} systematically examined the bug reports with smartphone vendors HTC and Motorola, providing evidence to point out the fragmentation issues in the Android ecosystem. Other related works~\cite{kamran2016android, mutchler2016target, wu2013impact, zhou2014peril} also revealed how severe the Android ecosystem has been suffering from fragmentation problems. Kamran et al.~\cite{kamran2016android} systematically classify Android fragmentation problems and discuss the solutions for developers to handle them.
Mutchler et al.~\cite{mutchler2016target} show that apps targeting
outdated Android versions would cause serious security consequences, such as incompatibility. Wu et al.~\cite{wu2013impact} investigated how vendor customizations impact overall Android security by analyzing representative stock Android images. Zhang et al.~\cite{zhang2019look} studied the compatibility intentions of Android apps from developers' perspective, revealing that malware developers' compatibility intentions were significantly different from those of benign apps. Zhou et al.~\cite{zhou2014peril} developed a tool for automatically detecting the security risks that lie in customized Android devices. 

Also, fragmentation can cause severe compatibility issues due to the API's fast-evolving~\cite{li2017static, li2018cid, liu2014characterizing, wei2016taming}. Liu et al.~\cite{liu2014characterizing} conducted an empirical study on performance bugs and summarize their common patterns, which reveals that performance bugs could be found on certain Android devices. In addition, Nayebi et al.~\cite{nayebi2012state} found that varying display resolutions of mobile devices are a serious challenge in Android development, leading to compatibility issues. These works demonstrate Android fragmentation is a major cause of Android security issues. The prevalence of such fragmentation issues motivated us to automatically detect them through crowdsourced testing.

\textbf{Detecting Android compatibility issues.}
To resolve the Android fragmentation issues, researchers leverage several testing tools~\cite{ham2014designing, kaasila2012testdroid, zhang2015compatibility,vilkomir2018multi, cheng2015mobile, naith2018hybrid, liu2016concurrent, lanui2019cloud, liu2019compatibility} for mobile application systems. For example, Ham et al.~\cite{ham2014designing} designed and implemented a compatibility testing system on top of the code level and the API level to handle fragmentation. Kaasila et al.~\cite{kaasila2012testdroid} presented Testdroid for conducting user interface tests on a variety of Android devices to identify crashes caused by fragmentation. Halpern et al.~\cite{halpern2015mosaic} proposed Mosaic, a record and replay tool to tame fragmentation through testing Android devices with different models.
Huang et al.~\cite{huang2014appacts} presented a Mobile App Automated Compatibility Testing Service (AppACTS), aiming at helping developers to conduct tests on devices more effectively. Ki et al.~\cite{ki2019mimic} further proposed a UI compatibility testing system, Mimic, that supports parallel testing across different Android or app versions.
Also, Google provides the Compatibility Test Suite (CTS) \cite{CTS} to tame fragmentation on Android devices. Also, Kong et al.~\cite{kong2018automated} conducted a systematic literature review on the state-of-the-art works of Android testing and concluded that taming Android ecosystem fragmentation is one of the most concrete research directions.
However, none of these works can be used in practical terms to execute test cases across all real-world Android devices. Apart from that, the state-of-the-art tool, CTS, is annoying users when running tests, and given there are 24,093~\cite{opensignal_report} distinct devices out there, it is nearly impossible for developers to collect all sorts of Android devices to identify compatibility issues. Apart from that, Li et al.~\cite{li2018cid} systematically modelled the life-cycle of the Android APIs to detect compatibility issues. In addition, Mahmud et al.~\cite{mahmud2021android} further proposed ACID that statically detected both API invocation compatibility issues and API callback
compatibility issues. Moreover, Silva et al.~\cite{silva2022saintdroid} introduced novel algorithms that automatically detect API and permission-induced incompatibilities. However, these static analysis approaches are known to suffer from imprecision when extracting the usage of the APIs, leading to lots of false-positive and false-negative results. To that end, \tool{} applies the dynamical technique to execute test cases on real-world devices to provide additional evidence in detecting compatibility issues that are overlooked by the state-of-the-art.

\textbf{Crowdsourced testing in Android.} 
Several techniques~\cite{wu2017appcheck, guo2020crowdsourced, li2019cocotest, liang2019summarizing, chen2019automatic, zhang2017crowdsourced} were proposed to do crowdsourced testing for android applications. Wu et al.~\cite{wu2017appcheck} proposed AppCheck, a crowdsourced testing service to automatically record the user interactions over the internet and then replayed them on real-world devices in order to identify compatibility issues. Li et al.~\cite{li2019cocotest}  developed a platform, CoCoTest, generating bug reports and recommending them to the other workers in real-time. Further, Zhang et al.~\cite{zhang2017crowdsourced} summarize informative concepts, insights, and challenges about common questions on crowdsourced test services. It indicates the challenges in performing large-scale user-oriented testing across Android devices, which this paper mainly focused on.

\section{CONCLUSIONS AND FUTURE WORK}
In this work, we presented a novel and lightweight prototype platform, \tool{}, that leverages crowdsourced testing techniques for pinpointing compatibility issues caused by Android fragmentation. Experimental results of thousands of test cases on a wide range of real-world Android devices show that (1) \tool{} is capable of automatically executing test cases on real-world Android devices in crowdsourced dispatching strategies;(2) Our approach is practical in automatically pinpointing and confirming API-induced compatibility issues. It also goes beyond the state-of-the-art to pinpoint not only signature-based compatibility issues but also semantics-based and vendor/model-based compatibility issues; (3) By investigating the customer experiences through qualitative metrics, users' satisfaction provides strong evidence that \tool{} is useful and welcome in practice.

In future work, we plan to explore the possibility of applying lightweight crowdsourced testing for more generic purposes going beyond taming Android fragmentation. Indeed, our approach, although proposed for crowdsourced testing for compatibility issues detection, is not strongly attached to identifying fragmentation issues. We believe it could be easily adapted to analyze other problems, such as security-related issues (e.g., to automatically detect hackable vulnerabilities by applying customized test cases). 

\section*{Acknowledgements}
The authors would like to thank the anonymous reviewers who have provided insightful and constructive comments on this paper. This work was supported by the Australian Research Council (ARC) under a Laureate Fellowship project FL190100035.

\balance

\bibliographystyle{IEEEtran}
\bibliography{main}

\begin{thebibliography}{10}
\providecommand{\url}[1]{#1}
\csname url@samestyle\endcsname
\providecommand{\newblock}{\relax}
\providecommand{\bibinfo}[2]{#2}
\providecommand{\BIBentrySTDinterwordspacing}{\spaceskip=0pt\relax}
\providecommand{\BIBentryALTinterwordstretchfactor}{4}
\providecommand{\BIBentryALTinterwordspacing}{\spaceskip=\fontdimen2\font plus
\BIBentryALTinterwordstretchfactor\fontdimen3\font minus
  \fontdimen4\font\relax}
\providecommand{\BIBforeignlanguage}[2]{{%
\expandafter\ifx\csname l@#1\endcsname\relax
\typeout{** WARNING: IEEEtran.bst: No hyphenation pattern has been}%
\typeout{** loaded for the language `#1'. Using the pattern for}%
\typeout{** the default language instead.}%
\else
\language=\csname l@#1\endcsname
\fi
#2}}
\providecommand{\BIBdecl}{\relax}
\BIBdecl

\bibitem{joorabchi2013real}
M.~E. Joorabchi, A.~Mesbah, and P.~Kruchten, ``Real challenges in mobile app
  development,'' in \emph{2013 ACM/IEEE International Symposium on Empirical
  Software Engineering and Measurement}.\hskip 1em plus 0.5em minus 0.4em\relax
  IEEE, 2013, pp. 15--24.

\bibitem{theverge_android}
{Russell Brandom}, ``{There are now 2.5 billion active Android devices},''
  \url{https://www.theverge.com/2019/5/7/18528297/google-io-2019-android-devices-play-store-total-number-statistic-keynote},
  online; accessed 15 Aug 2022.

\bibitem{liu2022customized}
P.~Liu, M.~Fazzini, J.~Grundy, and L.~Li, ``Do customized android frameworks
  keep pace with android?'' \emph{arXiv preprint arXiv:2205.15535}, 2022.

\bibitem{cai2019large}
H.~Cai, Z.~Zhang, L.~Li, and X.~Fu, ``A large-scale study of application
  incompatibilities in android,'' in \emph{Proceedings of the 28th ACM SIGSOFT
  International Symposium on Software Testing and Analysis}, 2019, pp.
  216--227.

\bibitem{he2018understanding}
D.~He, L.~Li, L.~Wang, H.~Zheng, G.~Li, and J.~Xue, ``Understanding and
  detecting evolution-induced compatibility issues in android apps,'' in
  \emph{2018 33rd IEEE/ACM International Conference on Automated Software
  Engineering (ASE)}.\hskip 1em plus 0.5em minus 0.4em\relax IEEE, 2018, pp.
  167--177.

\bibitem{scalabrino2019data}
S.~Scalabrino, G.~Bavota, M.~Linares-V{\'a}squez, M.~Lanza, and R.~Oliveto,
  ``Data-driven solutions to detect api compatibility issues in android: an
  empirical study,'' in \emph{2019 IEEE/ACM 16th International Conference on
  Mining Software Repositories (MSR)}.\hskip 1em plus 0.5em minus 0.4em\relax
  IEEE, 2019, pp. 288--298.

\bibitem{wei2016taming}
L.~Wei, Y.~Liu, and S.-C. Cheung, ``Taming android fragmentation:
  Characterizing and detecting compatibility issues for android apps,'' in
  \emph{Proceedings of the 31st IEEE/ACM International Conference on Automated
  Software Engineering}, 2016, pp. 226--237.

\bibitem{xia2020android}
H.~Xia, Y.~Zhang, Y.~Zhou, X.~Chen, Y.~Wang, X.~Zhang, S.~Cui, G.~Hong,
  X.~Zhang, M.~Yang \emph{et~al.}, ``How android developers handle
  evolution-induced api compatibility issues: a large-scale study,'' in
  \emph{2020 IEEE/ACM 42nd International Conference on Software Engineering
  (ICSE)}.\hskip 1em plus 0.5em minus 0.4em\relax IEEE, 2020, pp. 886--898.

\bibitem{ham2011mobile}
H.~K. Ham and Y.~B. Park, ``Mobile application compatibility test system design
  for android fragmentation,'' in \emph{International Conference on Advanced
  Software Engineering and Its Applications}.\hskip 1em plus 0.5em minus
  0.4em\relax Springer, 2011, pp. 314--320.

\bibitem{huang2018understanding}
H.~Huang, L.~Wei, Y.~Liu, and S.-C. Cheung, ``Understanding and detecting
  callback compatibility issues for android applications,'' in
  \emph{Proceedings of the 33rd ACM/IEEE International Conference on Automated
  Software Engineering}, 2018, pp. 532--542.

\bibitem{li2018cid}
L.~Li, T.~F. Bissyand{\'e}, H.~Wang, and J.~Klein, ``Cid: Automating the
  detection of api-related compatibility issues in android apps,'' in
  \emph{Proceedings of the 27th ACM SIGSOFT International Symposium on Software
  Testing and Analysis}, 2018, pp. 153--163.

\bibitem{zhang2015compatibility}
T.~Zhang, J.~Gao, J.~Cheng, and T.~Uehara, ``Compatibility testing service for
  mobile applications,'' in \emph{2015 IEEE Symposium on Service-Oriented
  System Engineering}.\hskip 1em plus 0.5em minus 0.4em\relax IEEE, 2015, pp.
  179--186.

\bibitem{sun2021taming}
X.~Sun, L.~Li, T.~F. Bissyand{\'e}, J.~Klein, D.~Octeau, and J.~Grundy,
  ``Taming reflection: An essential step toward whole-program analysis of
  android apps,'' \emph{ACM Transactions on Software Engineering and
  Methodology (TOSEM)}, vol.~30, no.~3, pp. 1--36, 2021.

\bibitem{liu2021identifying}
P.~Liu, L.~Li, Y.~Yan, M.~Fazzini, and J.~Grundy, ``Identifying and
  characterizing silently-evolved methods in the android api,'' in \emph{2021
  IEEE/ACM 43rd International Conference on Software Engineering: Software
  Engineering in Practice (ICSE-SEIP)}.\hskip 1em plus 0.5em minus 0.4em\relax
  IEEE, 2021, pp. 308--317.

\bibitem{sun2022mining}
X.~Sun, X.~Chen, Y.~Zhao, P.~Liu, J.~Grundy, and L.~Li, ``Mining android api
  usage to generate unit test cases for pinpointing compatibility issues,'' in
  \emph{37th IEEE/ACM International Conference on Automated Software
  Engineering}, 2022, pp. 1--13.

\bibitem{lu2016prada}
X.~Lu, X.~Liu, H.~Li, T.~Xie, Q.~Mei, D.~Hao, G.~Huang, and F.~Feng, ``Prada:
  Prioritizing android devices for apps by mining large-scale usage data,'' in
  \emph{2016 IEEE/ACM 38th International Conference on Software Engineering
  (ICSE)}.\hskip 1em plus 0.5em minus 0.4em\relax IEEE, 2016, pp. 3--13.

\bibitem{choudhary2015automated}
S.~R. Choudhary, A.~Gorla, and A.~Orso, ``Automated test input generation for
  android: Are we there yet?(e),'' in \emph{2015 30th IEEE/ACM International
  Conference on Automated Software Engineering (ASE)}.\hskip 1em plus 0.5em
  minus 0.4em\relax IEEE, 2015, pp. 429--440.

\bibitem{han2012understanding}
D.~Han, C.~Zhang, X.~Fan, A.~Hindle, K.~Wong, and E.~Stroulia, ``Understanding
  android fragmentation with topic analysis of vendor-specific bugs,'' in
  \emph{2012 19th Working Conference on Reverse Engineering}.\hskip 1em plus
  0.5em minus 0.4em\relax IEEE, 2012, pp. 83--92.

\bibitem{wei2018understanding}
L.~Wei, Y.~Liu, S.-C. Cheung, H.~Huang, X.~Lu, and X.~Liu, ``Understanding and
  detecting fragmentation-induced compatibility issues for android apps,''
  \emph{IEEE Transactions on Software Engineering}, vol.~46, no.~11, pp.
  1176--1199, 2018.

\bibitem{CTS}
\BIBentryALTinterwordspacing
\emph{Compatibility Test Suite}, 2021. [Online]. Available:
  \url{https://source.android.com/compatibility/cts}
\BIBentrySTDinterwordspacing

\bibitem{park2013fragmentation}
J.-H. Park, Y.~B. Park, and H.~K. Ham, ``Fragmentation problem in android,'' in
  \emph{2013 International Conference on Information Science and Applications
  (ICISA)}.\hskip 1em plus 0.5em minus 0.4em\relax IEEE, 2013, pp. 1--2.

\bibitem{liu2016concurrent}
C.-H. Liu, W.-K. Chen, and S.-L. Chen, ``A concurrent approach for improving
  the efficiency of android cts testing,'' in \emph{2016 International Computer
  Symposium (ICS)}.\hskip 1em plus 0.5em minus 0.4em\relax IEEE, 2016, pp.
  611--615.

\bibitem{zhou2014peril}
X.~Zhou, Y.~Lee, N.~Zhang, M.~Naveed, and X.~Wang, ``The peril of
  fragmentation: Security hazards in android device driver customizations,'' in
  \emph{2014 IEEE Symposium on Security and Privacy}.\hskip 1em plus 0.5em
  minus 0.4em\relax IEEE, 2014, pp. 409--423.

\bibitem{wu2013impact}
L.~Wu, M.~Grace, Y.~Zhou, C.~Wu, and X.~Jiang, ``The impact of vendor
  customizations on android security,'' in \emph{Proceedings of the 2013 ACM
  SIGSAC conference on Computer \& communications security}, 2013, pp.
  623--634.

\bibitem{orso2002gamma}
A.~Orso, D.~Liang, M.~J. Harrold, and R.~Lipton, ``Gamma system: Continuous
  evolution of software after deployment,'' in \emph{Proceedings of the 2002
  ACM SIGSOFT international symposium on Software testing and analysis}, 2002,
  pp. 65--69.

\bibitem{memon2004skoll}
A.~Memon, A.~Porter, C.~Yilmaz, A.~Nagarajan, D.~Schmidt, and B.~Natarajan,
  ``Skoll: Distributed continuous quality assurance,'' in \emph{Proceedings.
  26th International Conference on Software Engineering}.\hskip 1em plus 0.5em
  minus 0.4em\relax IEEE, 2004, pp. 459--468.

\bibitem{elbaum2004empirical}
S.~Elbaum and M.~Hardojo, ``An empirical study of profiling strategies for
  released software and their impact on testing activities,'' in
  \emph{Proceedings of the 2004 ACM SIGSOFT international symposium on Software
  testing and analysis}, 2004, pp. 65--75.

\bibitem{Global_App_Testing}
\BIBentryALTinterwordspacing
\emph{Global App Testing}, 2022. [Online]. Available:
  \url{https://go.globalapptesting.com/app-testing-for-engineering-qa}
\BIBentrySTDinterwordspacing

\bibitem{Digivante}
\BIBentryALTinterwordspacing
\emph{Digivante}, 2022. [Online]. Available:
  \url{https://www.digivante.com/crowdsourced-testing-referral/?utm_campaign=SoftwareTestingHelp\%20Referral\%20Campaigns&utm_source=software-testing-help&utm_content=crowdtesting}
\BIBentrySTDinterwordspacing

\bibitem{test_IO}
\BIBentryALTinterwordspacing
\emph{test IO}, 2022. [Online]. Available: \url{https://goo.gl/rGQPWF}
\BIBentrySTDinterwordspacing

\bibitem{QA_Mentor}
\BIBentryALTinterwordspacing
\emph{QA Mentor}, 2022. [Online]. Available:
  \url{https://www.qamentor.com/qa-services/crowdsourced-testing-services/}
\BIBentrySTDinterwordspacing

\bibitem{wu2017appcheck}
G.~Wu, Y.~Cao, W.~Chen, J.~Wei, H.~Zhong, and T.~Huang, ``Appcheck: a
  crowdsourced testing service for android applications,'' in \emph{2017 IEEE
  International Conference on Web Services (ICWS)}.\hskip 1em plus 0.5em minus
  0.4em\relax IEEE, 2017, pp. 253--260.

\bibitem{guo2020crowdsourced}
C.~Guo, T.~He, W.~Yuan, Y.~Guo, and R.~Hao, ``Crowdsourced requirements
  generation for automatic testing via knowledge graph,'' in \emph{Proceedings
  of the 29th ACM SIGSOFT International Symposium on Software Testing and
  Analysis}, 2020, pp. 545--548.

\bibitem{li2019cocotest}
H.~Li, C.~Fang, Z.~Wei, and Z.~Chen, ``Cocotest: collaborative crowdsourced
  testing for android applications,'' in \emph{Proceedings of the 28th ACM
  SIGSOFT International Symposium on Software Testing and Analysis}, 2019, pp.
  390--393.

\bibitem{liang2019summarizing}
H.~Liang and T.~He, ``Summarizing the crowdsourced testing,'' in \emph{2019
  IEEE 5th International Conference on Computer and Communications
  (ICCC)}.\hskip 1em plus 0.5em minus 0.4em\relax IEEE, 2019, pp. 526--530.

\bibitem{chen2019automatic}
X.~Chen, H.~Jiang, Z.~Chen, T.~He, and L.~Nie, ``Automatic test report
  augmentation to assist crowdsourced testing,'' \emph{Frontiers of Computer
  Science}, vol.~13, no.~5, pp. 943--959, 2019.

\bibitem{zhang2017crowdsourced}
T.~Zhang, J.~Gao, and J.~Cheng, ``Crowdsourced testing services for mobile
  apps,'' in \emph{2017 IEEE Symposium on Service-Oriented System Engineering
  (SOSE)}.\hskip 1em plus 0.5em minus 0.4em\relax IEEE, 2017, pp. 75--80.

\bibitem{wei2019pivot}
L.~Wei, Y.~Liu, and S.-C. Cheung, ``Pivot: learning api-device correlations to
  facilitate android compatibility issue detection,'' in \emph{2019 IEEE/ACM
  41st International Conference on Software Engineering (ICSE)}.\hskip 1em plus
  0.5em minus 0.4em\relax IEEE, 2019, pp. 878--888.

\bibitem{bao2019android}
J.~Bao, \emph{Android App-Hook and Plug-In Technology}.\hskip 1em plus 0.5em
  minus 0.4em\relax CRC Press, 2019.

\bibitem{tinker}
\BIBentryALTinterwordspacing
\emph{Tinker}, 2021. [Online]. Available:
  \url{https://github.com/Tencent/tinker}
\BIBentrySTDinterwordspacing

\bibitem{JUnit}
\BIBentryALTinterwordspacing
\emph{JUnit}, 2021. [Online]. Available:
  \url{https://en.wikipedia.org/wiki/JUnit#:~:text=JUnit%20is%20a%20unit%20testing,xUnit%20that%20originated%20with%20SUnit.&text=junit%20and%20junit.}
\BIBentrySTDinterwordspacing

\bibitem{sourcecode_aosp}
\BIBentryALTinterwordspacing
\emph{Source Code of the Android Open Source Project}, 2021. [Online].
  Available: \url{https://cs.android.com/android}
\BIBentrySTDinterwordspacing

\bibitem{fraser2011evosuite}
G.~Fraser and A.~Arcuri, ``Evosuite: automatic test suite generation for
  object-oriented software,'' in \emph{Proceedings of the 19th ACM SIGSOFT
  symposium and the 13th European conference on Foundations of software
  engineering}, 2011, pp. 416--419.

\bibitem{hotswap}
{wikipedia}, ``{Hot swapping},''
  \url{https://en.wikipedia.org/wiki/Hot_swapping}, online; accessed 28 January
  2022.

\bibitem{liu2020androzooopen}
P.~Liu, L.~Li, Y.~Zhao, X.~Sun, and J.~Grundy, ``Androzooopen: Collecting
  large-scale open source android apps for the research community,'' in
  \emph{Proceedings of the 17th International Conference on Mining Software
  Repositories}, 2020, pp. 548--552.

\bibitem{stackoverflowSamsung}
{Stack OverFlow}, ``{Native error on Android 8.0 Samsung S8},''
  \url{https://stackoverflow.com/questions/50360227/tgkill-native-error-on-android-8-0-samsung-s8},
  online; accessed 02 December 2021.

\bibitem{Githubhiddenapi}
{Github}, ``{android-hidden-api},''
  \url{https://github.com/anggrayudi/android-hidden-api}, online; accessed 02
  December 2021.

\bibitem{li2016accessing}
L.~Li, T.~F. Bissyand{\'e}, Y.~Le~Traon, and J.~Klein, ``Accessing inaccessible
  android apis: An empirical study,'' in \emph{2016 IEEE International
  Conference on Software Maintenance and Evolution (ICSME)}.\hskip 1em plus
  0.5em minus 0.4em\relax IEEE, 2016, pp. 411--422.

\bibitem{Likertscale}
{wikipedia}, ``{Likert scale},''
  \url{https://en.wikipedia.org/wiki/Likert_scale}, online; accessed 13 January
  2022.

\bibitem{CTS_setup}
\BIBentryALTinterwordspacing
\emph{CTS setup steps}, 2021. [Online]. Available:
  \url{https://source.android.com/compatibility/cts/setup}
\BIBentrySTDinterwordspacing

\bibitem{marketshare}
{Statcounter}, ``{Mobile Vendor Market Share Worldwide},''
  \url{https://gs.statcounter.com/vendor-market-share/mobile}, online; accessed
  02 December 2021.

\bibitem{gao2019successes}
R.~Gao, Y.~Wang, Y.~Feng, Z.~Chen, and W.~E. Wong, ``Successes, challenges, and
  rethinking--an industrial investigation on crowdsourced mobile application
  testing,'' \emph{Empirical Software Engineering}, vol.~24, no.~2, pp.
  537--561, 2019.

\bibitem{mahmud2022android}
T.~Mahmud, M.~Che, and G.~Yang, ``Android api field evolution and its induced
  compatibility issues,'' in \emph{Proceedings of the 16th ACM/IEEE
  International Symposium on Empirical Software Engineering and Measurement},
  2022, pp. 34--44.

\bibitem{mahmud2021android}
{Mahmud, Tarek and Che, Meiru and Yang, Guowei}, ``Android compatibility issue
  detection using api differences,'' in \emph{2021 IEEE International
  Conference on Software Analysis, Evolution and Reengineering (SANER)}.\hskip
  1em plus 0.5em minus 0.4em\relax IEEE, 2021, pp. 480--490.

\bibitem{liu2022automatically}
P.~Liu, Y.~Zhao, H.~Cai, M.~Fazzini, J.~Grundy, and L.~Li, ``Automatically
  detecting api-induced compatibility issues in android apps: A comparative
  analysis (replicability study),'' \emph{arXiv preprint arXiv:2205.15561},
  2022.

\bibitem{cotroneo2020comprehensive}
D.~Cotroneo, A.~K. Iannillo, R.~Natella, and R.~Pietrantuono, ``A comprehensive
  study on software aging across android versions and vendors,''
  \emph{Empirical Software Engineering}, vol.~25, no.~5, pp. 3357--3395, 2020.

\bibitem{li2018elegant}
C.~Li, C.~Xu, L.~Wei, J.~Wang, J.~Ma, and J.~Lu, ``Elegant: Towards effective
  location of fragmentation-induced compatibility issues for android apps,'' in
  \emph{2018 25th Asia-Pacific Software Engineering Conference (APSEC)}.\hskip
  1em plus 0.5em minus 0.4em\relax IEEE, 2018, pp. 278--287.

\bibitem{huang2021characterizing}
H.~Huang, M.~Wen, L.~Wei, Y.~Liu, and S.-C. Cheung, ``Characterizing and
  detecting configuration compatibility issues in android apps,'' \emph{arXiv
  preprint arXiv:2109.00300}, 2021.

\bibitem{kamran2016android}
M.~Kamran, J.~Rashid, and M.~W. Nisar, ``Android fragmentation classification,
  causes, problems and solutions,'' \emph{International Journal of Computer
  Science and Information Security}, vol.~14, no.~9, p. 992, 2016.

\bibitem{mutchler2016target}
P.~Mutchler, Y.~Safaei, A.~Doup{\'e}, and J.~Mitchell, ``Target fragmentation
  in android apps,'' in \emph{2016 IEEE Security and Privacy Workshops
  (SPW)}.\hskip 1em plus 0.5em minus 0.4em\relax IEEE, 2016, pp. 204--213.

\bibitem{zhang2019look}
Z.~Zhang and H.~Cai, ``A look into developer intentions for app compatibility
  in android,'' in \emph{2019 IEEE/ACM 6th International Conference on Mobile
  Software Engineering and Systems (MOBILESoft)}.\hskip 1em plus 0.5em minus
  0.4em\relax IEEE, 2019, pp. 40--44.

\bibitem{li2017static}
L.~Li, T.~F. Bissyand{\'e}, M.~Papadakis, S.~Rasthofer, A.~Bartel, D.~Octeau,
  J.~Klein, and L.~Traon, ``Static analysis of android apps: A systematic
  literature review,'' \emph{Information and Software Technology}, vol.~88, pp.
  67--95, 2017.

\bibitem{liu2014characterizing}
Y.~Liu, C.~Xu, and S.-C. Cheung, ``Characterizing and detecting performance
  bugs for smartphone applications,'' in \emph{Proceedings of the 36th
  international conference on software engineering}, 2014, pp. 1013--1024.

\bibitem{nayebi2012state}
F.~Nayebi, J.-M. Desharnais, and A.~Abran, ``The state of the art of mobile
  application usability evaluation,'' in \emph{2012 25th IEEE Canadian
  Conference on Electrical and Computer Engineering (CCECE)}.\hskip 1em plus
  0.5em minus 0.4em\relax IEEE, 2012, pp. 1--4.

\bibitem{ham2014designing}
H.~K. Ham and Y.~B. Park, ``Designing knowledge base mobile application
  compatibility test system for android fragmentation,'' \emph{International
  Journal of Software Engineering and Its Applications}, vol.~8, no.~1, pp.
  303--314, 2014.

\bibitem{kaasila2012testdroid}
J.~Kaasila, D.~Ferreira, V.~Kostakos, and T.~Ojala, ``Testdroid: automated
  remote ui testing on android,'' in \emph{Proceedings of the 11th
  International Conference on Mobile and Ubiquitous Multimedia}, 2012, pp.
  1--4.

\bibitem{vilkomir2018multi}
S.~Vilkomir, ``Multi-device coverage testing of mobile applications,''
  \emph{Software quality journal}, vol.~26, no.~2, pp. 197--215, 2018.

\bibitem{cheng2015mobile}
J.~Cheng, Y.~Zhu, T.~Zhang, C.~Zhu, and W.~Zhou, ``Mobile compatibility testing
  using multi-objective genetic algorithm,'' in \emph{2015 IEEE Symposium on
  Service-Oriented System Engineering}.\hskip 1em plus 0.5em minus 0.4em\relax
  IEEE, 2015, pp. 302--307.

\bibitem{naith2018hybrid}
Q.~Naith and F.~Ciravegna, ``Hybrid crowd-powered approach for compatibility
  testing of mobile devices and applications,'' in \emph{Proceedings of the 3rd
  International Conference on Crowd Science and Engineering}, 2018, pp. 1--8.

\bibitem{lanui2019cloud}
A.~Lanui and T.~K. Chiew, ``A cloud-based solution for testing applications'
  compatibility and portability on fragmented android platform,'' in \emph{2019
  26th Asia-Pacific Software Engineering Conference (APSEC)}.\hskip 1em plus
  0.5em minus 0.4em\relax IEEE, 2019, pp. 158--164.

\bibitem{liu2019compatibility}
C.-H. Liu, ``A compatibility testing platform for android multimedia
  applications,'' \emph{Multimedia Tools and Applications}, vol.~78, no.~4, pp.
  4885--4904, 2019.

\bibitem{halpern2015mosaic}
M.~Halpern, Y.~Zhu, R.~Peri, and V.~J. Reddi, ``Mosaic: cross-platform
  user-interaction record and replay for the fragmented android ecosystem,'' in
  \emph{2015 IEEE International Symposium on Performance Analysis of Systems
  and Software (ISPASS)}.\hskip 1em plus 0.5em minus 0.4em\relax IEEE, 2015,
  pp. 215--224.

\bibitem{huang2014appacts}
J.-f. Huang, ``Appacts: Mobile app automated compatibility testing service,''
  in \emph{2014 2nd IEEE International Conference on Mobile Cloud Computing,
  Services, and Engineering}.\hskip 1em plus 0.5em minus 0.4em\relax IEEE,
  2014, pp. 85--90.

\bibitem{ki2019mimic}
T.~Ki, C.~M. Park, K.~Dantu, S.~Y. Ko, and L.~Ziarek, ``Mimic: Ui compatibility
  testing system for android apps,'' in \emph{2019 IEEE/ACM 41st International
  Conference on Software Engineering (ICSE)}.\hskip 1em plus 0.5em minus
  0.4em\relax IEEE, 2019, pp. 246--256.

\bibitem{kong2018automated}
P.~Kong, L.~Li, J.~Gao, K.~Liu, T.~F. Bissyand{\'e}, and J.~Klein, ``Automated
  testing of android apps: A systematic literature review,'' \emph{IEEE
  Transactions on Reliability}, vol.~68, no.~1, pp. 45--66, 2018.

\bibitem{opensignal_report}
\BIBentryALTinterwordspacing
\emph{Android fragmentation report}, 2015. [Online]. Available:
  \url{https://www.opensignal.com/sites/opensignal-com/files/data/reports/global/data-2015-08/2015_08_fragmentation_report.pdf}
\BIBentrySTDinterwordspacing

\bibitem{silva2022saintdroid}
B.~Silva, C.~Stevens, N.~Mansoor, W.~Srisa-An, T.~Yu, and H.~Bagheri,
  ``Saintdroid: Scalable, automated incompatibility detection for android,'' in
  \emph{2022 52nd Annual IEEE/IFIP International Conference on Dependable
  Systems and Networks (DSN)}.\hskip 1em plus 0.5em minus 0.4em\relax IEEE,
  2022, pp. 567--579.

\end{thebibliography}

\end{document}